\documentclass[twocolumn]{aa}
\newcommand{\II}{\scriptsize{II}\normalsize}

\usepackage{natbib}
\usepackage{graphicx}
\usepackage{multirow}
\usepackage{lscape}

\usepackage{txfonts}
\usepackage{soul}
\def\d{$^\circ$}
\def\m{$^\prime$}
\def\s{$^{\prime\prime}$}
\def\hh{$^{\mathrm h}$}
\def\mm{$^{\mathrm m}$}
\def\ss{$^{\mathrm s}$}

\usepackage{amstext}

\begin{document}

   \title{Activity-rotation in the dM4 star Gl 729.}

   \subtitle{A possible chromospheric cycle.}

   \author{R.V. Iba\~nez Bustos
\inst{1,2}\fnmsep\thanks{Contact e-mail: ribanez@iafe.uba.ar}
          \thanks{Based on data obtained at Complejo Astron\'omico El Leoncito, operated under agreement between the Consejo Nacional de Investigaciones Cient\'\i ficas y T\'ecnicas de la Rep\'ublica Argentina and the National Universities of La Plata, C\'ordoba and San Juan.}
          \and
          A.P. Buccino\inst{1,3}
          \and
          S. Messina\inst{4}
          \and
          A.F. Lanza\inst{4}
          \and
          P.J.D. Mauas\inst{1,3}
          }

   \institute{Instituto de Astronom\'ia y F\'isica del Espacio (CONICET-UBA), C.C. 67 Sucursal 28, C1428EHA-Buenos Aires, Argentina.\\
        \and
             Departamento de F\'isica. FI-Universidad de Buenos Aires, Buenos Aires, Argentina.\\
        \and
             Departamento de F\'isica. FCEyN-Universidad de Buenos Aires, Buenos Aires, Argentina.\\
        \and
             INAF-Osservatorio Astrofisico di Catania, via S. Sofia 78, 95123, Catania, Italia. 
            }

   \date{}

% \abstract{}{}{}{}{} 
% 5 {} token are mandatory
 
  \abstract 
    {}
{Recently, new debates about the  role of layers  of strong shear have emerged in stellar dynamo  theory. Further information on the long-term magnetic activity of fully convective stars could help determine whether their underlying dynamo could sustain activity cycles similar to the solar one. }
{We performed  a thorough study of  the short- and long-term magnetic activity of the young active dM4 star Gl 729. First, we analyzed long-cadence $K2$ photometry to characterize its transient events (e.g., flares) and global and surface differential rotation.
Then, from the Mount Wilson $S$-indexes derived from CASLEO spectra and other public observations, we analyzed its long-term activity between 1998  and 2020  with four different time-domain techniques to detect cyclic patterns.
Finally, we explored the chromospheric activity at different heights with simultaneous measurements of the H$\alpha$ and the Na I D indexes, and we analyzed their relations with the $S$-Index.
}
{We found that the cumulative flare frequency follows a power-law distribution  with slope $\sim- 0.73$ for the range $10^{32}$ to $10^{34}$ erg.
We obtained $P_{rot} = (2.848 \pm 0.001)$ days, and we found no evidence of differential rotation.
We also found that this young active star presents a long-term activity cycle with a length of $\text{about four}$ years; there is less significant evidence of a shorter cycle of $0.8$ year. The star also shows a broad activity minimum  between 1998 and 2004. 
We found a correlation between the S index, on the one hand, and the H$\alpha$ the Na I D indexes, on the other hand, although the saturation level of these last two indexes is not observed in the Ca lines.
} 
{Because the maximum-entropy spot model does not reflect migration between active longitudes, this activity cycle cannot be explained by a solar-type dynamo. It is probably caused by  an $\alpha^2$-dynamo.
}


   \maketitle
%
%________________________________________________________________

\section{Introduction}

M dwarfs, with masses between 0.1 and 0.5 M$_\odot$, constitute $\text{about } 75 \%$ of the stars in the solar neighborhood.   
Approximately half of them show high chromospheric activity that exceeds the activity of the Sun. They are usually referred to as dMe stars because they have H$\alpha$ in emission, or are called flare stars because they show frequent highly energetic flares (e.g., \citealt{Gunther20,Rodriguez20}). The fraction of these active stars depends markedly on the spectral class \citep{West04,West11,Reiners12}.  

Stellar activity is assumed to be driven by a stellar dynamo. For solar activity in particular, this is the $\alpha\Omega$ dynamo, which is explained by the feedback between the differential rotation of the star, which stretches the lines of the poloidal magnetic field to produce a toroidal field (the $\Omega$-effect), and the turbulent helical movements of the plasma in the convective zone, which regenerates the poloidal field (the $\alpha$-effect, see \cite{Charbonneau10} for more details of the solar dynamo process).
In several dynamo models  (e.g., \citealt{Charboneau97,Dikipati05}), the tachocline, an interface between the rigidly rotating radiative core and the convection zone, plays a fundamental role. A solar-type dynamo can also successfully reproduce  stellar activity cycles in cooler stars (e.g., \citealt{Buccino20}). 

However, dM stars with masses lower than $\sim$0.35 M$_\odot$ are thought to be fully convective \citep{Chabrier97} and therefore should not have a tachocline. Despite this fundamental difference in stellar structure, strong  magnetic activity is also observed in very low-mass fully convective stars (\citealt{Mauas96,Hawley96, West04}). Furthermore, stellar activity cycles were also detected in these stars (\citealt{Cincunegui07, Diaz07, SuarezMascareno16, Wargelin17,Ibanez19}).
A new debate  on the role of layers of strong shear has recently been initiated  by  \cite{Wright16}. 
They suggested that the presence of a tachocline is not a critical ingredient in a solar-type dynamo.  
Based on this, \cite{Yadav16} developed a  dynamo model for the fully convective star Proxima Centauri and concluded that large Rossby numbers (the ratio of stellar rotation period to convective turnover time) may promote regular activity cycles in fully convective stars. 
Therefore observational evidence of cyclic activity near the convection threshold and beyond can contribute new  insight into this discussion \citep{Buccino14,Ibanez19,Toledo19}.

The relation between stellar rotation and magnetic activity is another fundamental key to understanding the stellar dynamo. It shows an increase  in activity when the rotation period decreases, with a saturated activity regime for fast rotators. 
This relation has been widely studied in different works using different indicators: by \cite{Wright11}, \cite{Wright16}, and \cite{Wright18} in X-ray emission (using $L_X / L_{bol}$), and  by \cite{Astudillo17} and \cite{Newton17} in the optical range, employing the $\log(R'_{HK})$ and $L_{H\alpha} / L_{bol}$, respectively. 
In both the optical and X-ray ranges, these studies concluded that the transition in the inner structure of the stars does not imply a change in the dynamo mechanism (e.g., \citealt{Wright16}), and stellar rotation therefore plays an important role in driving the activity also in fully convective stars (e.g., \citealt{Ibanez19}). Even in this case, however, \cite{Astudillo17} found three stars that do not belong to either of these  regimes.

We study Gliese 729 in detail, which is one of the outliers in the saturation regime reported by \cite{Astudillo17}  because its $\log(R'_{HK})$ value deviates by more than 3$\sigma$ from the fit (see Fig. 6 in \citeauthor{Astudillo17}). 
We present a short- and long-term activity analysis of this star using high-precision photometry and a long series of spectroscopic observations.
In section \S\ref{sec.target} we describe some features that make Gl 729 a particularly interesting target. 
In section \S\ref{sec.obs} we provide an overview of our optical spectroscopic observations together with the archival data that we also employed for the analysis. 
In sections \S\ref{sec.res} and \S\ref{sec.spec} we present a detailed analysis of the photometric and spectroscopic data with different time-domain techniques, and we explore the relation between different spectral features at different activity levels. 
Finally, in section \S\ref{sec.concl} we discuss the main conclusions of this analysis.

%__________________________________________________________________
%__________________________________________________________________
%__________________________________________________________________

\section{The target} \label{sec.target}

Gl 729 (Ross 154, HIP 92403, V* V1216 Sgr) is an M4  active dwarf of the southern constellation of Sagittarius ($\alpha_{2000}$$=$18\hh 549 \mm 49.4 \ss, $\delta_{2000}$$=-$23\d50\m10\s). 
This single star is the seventh M dwarf closest to our Sun with a distance of 2.97 pc, a radius of 0.19 $R_{\odot}$ , and a mass of 0.14 $M_\odot$ \citep{Gaidos14}. 

Gl 729 is a flare star that has been observed in the optical (\citealt{Falchi90, Astudillo17}), X-ray (\citealt{Wargelin08, Malo14III}), and extreme-UV (EUV) ranges \citep{Tsikoudi97}.
First, \cite{Kiraga07} reported a rotation period of $P_{rot} = 2.869$ d  for Gl 729 using ASAS photometry, confirmed by \cite{DiezAlonso19}.
Based on this period and  its coronal activity level,  given by $\log (L_X/L_{bol}) = -3.5 $ \citep{Johns96}, Gl 729 is below  the saturation regime ($L_X/L_{bol} = 10^{-3}$) reported by \cite{Wright11}.

\cite{Wargelin08} studied the coronal magnetic activity using two  Chandra observations. 
One of them presents a very large flare with evidence of the Neupert effect. 
The other observation has several  moderate flares.
The authors found that the distribution of flare intensities does not appear to follow a single power law, as in the solar case.
They analyzed the nonflaring phase to search for low-level flaring, and found that the microflaring explains the emission in this ``quiescent'' regime. They found that the normalized X-ray luminosity $L_X/L_{bol}$ is below $10^{-3}$, which is near the mean value of the  saturation regime.

Because of their low masses, M dwarf stars are also ideal targets for detecting terrestrial planets from the ground (e.g., \citealt{Anglada16,Ribas18,Diaz19}) or from space (e.g., \citealt{Murihead12,Winters19}). 
Furthermore, statistical studies show a high occurrence of small planets around M stars \citep{Bonfils13,DressingCharbonneau15}. 
However, activity features could hide an exoplanet signal and even mimic it \citep{Robertson14}. 
Therefore, great efforts have been made to distinguish activity and planetary signals in  the radial velocity series (e.g., \citealt{Desort07,Haywood14,Diaz16}) or transiting curves (e.g., \citealt{Boisse12,Bonomo12,Aigrain16,Morris20}). 
For this reason, an exhaustive characterization of the stellar magnetic activity  over different timescales might allow detecting extrasolar planets even around active stars. 
In particular, Gl 729 belongs to the CARMENES (Calar Alto high-Resolution search for M dwarfs with Exoearths with Near-infrared and optical Échelle Spectrographs) input catalog \citep{Reiners18}, which is not biased by the activity level. 
Although Gl 729 has been observed by different programs, it has not been studied in the optical range in detail.
We present an extensive study of this dMe variable star based on almost 20 years of spectroscopic observations.

%__________________________________________________________________
%__________________________________________________________________
%__________________________________________________________________

\section{Observations}\label{sec.obs}

The HK$\alpha$ Project was started in 1999 with the aim to study the long-term chromospheric activity of southern cool stars.
In  this  program, we systematically observe late-type stars from dF5 to dM5.5 with the 2.15 m Jorge Sahade telescope at the CASLEO observatory, which is located at 2552 m above sea level in the Argentinian Andes.
The medium-resolution echelle spectra ($R\approx 13.000$) were obtained with the REOSC\footnote{\textsf{http://www.casleo.gov.ar/instrumental/js-reosc.php}} spectrograph.
They cover a maximum wavelength range between 3860 and 6690 \AA. 
We calibrate  all our echelle spectra in flux using IRAF\footnote{The Image Reduction and Analysis Facility (IRAF) is distributed by the Association of Universities for Research in Astronomy (AURA),  Inc., under contract to the National Science Foundation} routines 
and following the procedure described in \cite{Cincu04}.
We here employed 19 observations distributed between 2005 and 2019. Each observation consists of two consecutive spectra with 45 minutes exposure time each, which are combined to eliminate cosmic rays and to reduce noise. 

We complemented our data with public observations obtained with several spectrographs by the programs listed in Table \ref{obs_table}. 
Sixty echelle spectra were observed by HARPS, mounted at the 3.6 m telescope (\textit{R} $\sim$ 115.000) at La Silla Observatory (LSO, Chile), distributed over 2005 and the interval $2015-2017$.
Only one FEROS spectrum was available in our study. 
This spectrograph is placed on the 2.2 m telescope in LSO and has a resolution of $R \sim 48.000$.
Four spectra were taken in 2011 and 2015 with UVES, attached to the Unit Telescope 2 (UT2) of the Very Large Telescope (VLT) ($R \sim 80.000$) at Paranal Observatory.
Two medium-resolution spectra (\textit{R} $\sim$ 8.900) in the UVB wavelength range (300 - 559.5 nm) were obtained in 2014 with the X-SHOOTER spectrograph, mounted at the UT2 Cassegrain focus also at the VLT. 
Finally, we employed 13 spectra from the HIRES spectrograph mounted at the Keck-I telescope that were observed in 1998 and 2010.

HARPS and FEROS spectra have been automatically processed by their respective pipelines\footnote{\textsf{http://www.eso.org/sci/facilities/lasilla/instruments/harps.html}}$^,$\footnote{\textsf{http://www.eso.org/sci/facilities/lasilla/instruments/feros.html}}, while UVES and XSHOOTER observations  were manually calibrated with the corresponding procedure\footnote{\textsf{http://www.eso.org/sci/facilities/paranal/instruments/uves.html}}$^,$\footnote{\textsf{http://www.eso.org/sci/facilities/paranal/instruments/xshooter.html}}. We also analyzed high-precision photometry obtained during the \textit{K2} mission by the \textit{Kepler} spacecraft \citep{Borucki10}. 
\textit{K2} observed a total of 20 fields in sequential series of observational campaigns lasting $\sim$80 d each.  
Throughout the mission,  \textit{K2}  observed in  two  cadence  modes: short cadence (one observation every minute) and long  cadence (one observation every $\sim$30 minutes).  
In this study, we analyze  the long-cadence observations of Gl 729 during campaign 7 of the GO7016\_GO7060 proposal, obtained between 2015 October 4 and 2015 December 26. 

%__________________________________________________________________
%__________________________________________________________________
%__________________________________________________________________

\section{Analysis of $K2$ Photometry}\label{sec.res}

In this section we first present an analysis of the short-term activity of Gl 729, derived from the \textit{Kepler} light curve. 
We also model the stellar variability with a maximum-entropy spot model to derive active longitudes and estimate a minimum amplitude for the stellar differential rotation.

\subsection{Flare activity and rotation} \label{short-term}

To study the magnetic variability of Gl 729, we employed high-quality photometry obtained by the $K2$ mission.
The \textit{Kepler} database provides short- and long-cadence light curves, which constitute a great basis for  detecting flare-like events (e.g., \citealt{Hawley14,Davenport16}).  
Only long-cadence photometry is available for Gl 729.

In order to analyze both the rotational modulation and detect these transient events in the \textit{Kepler} light curve, we analyzed the time series with  the flare detection with ransac method
(FLATW'RM) algorithm  based on machine-learning techniques \citep{Vida18}\footnote{FLATW'RM is available at \textsf{https://github.com/vidakris/flatwrm/}.}. 
In particular, the FLATW'RM code first determines the stellar rotation period,  and after subtracting the  fitted rotational modulation from the light curve, it detects the flare-like events and reports the  starting and ending time, and the time of maximum flare flux. 

\begin{figure*}
\centering
   \includegraphics[width=\textwidth]{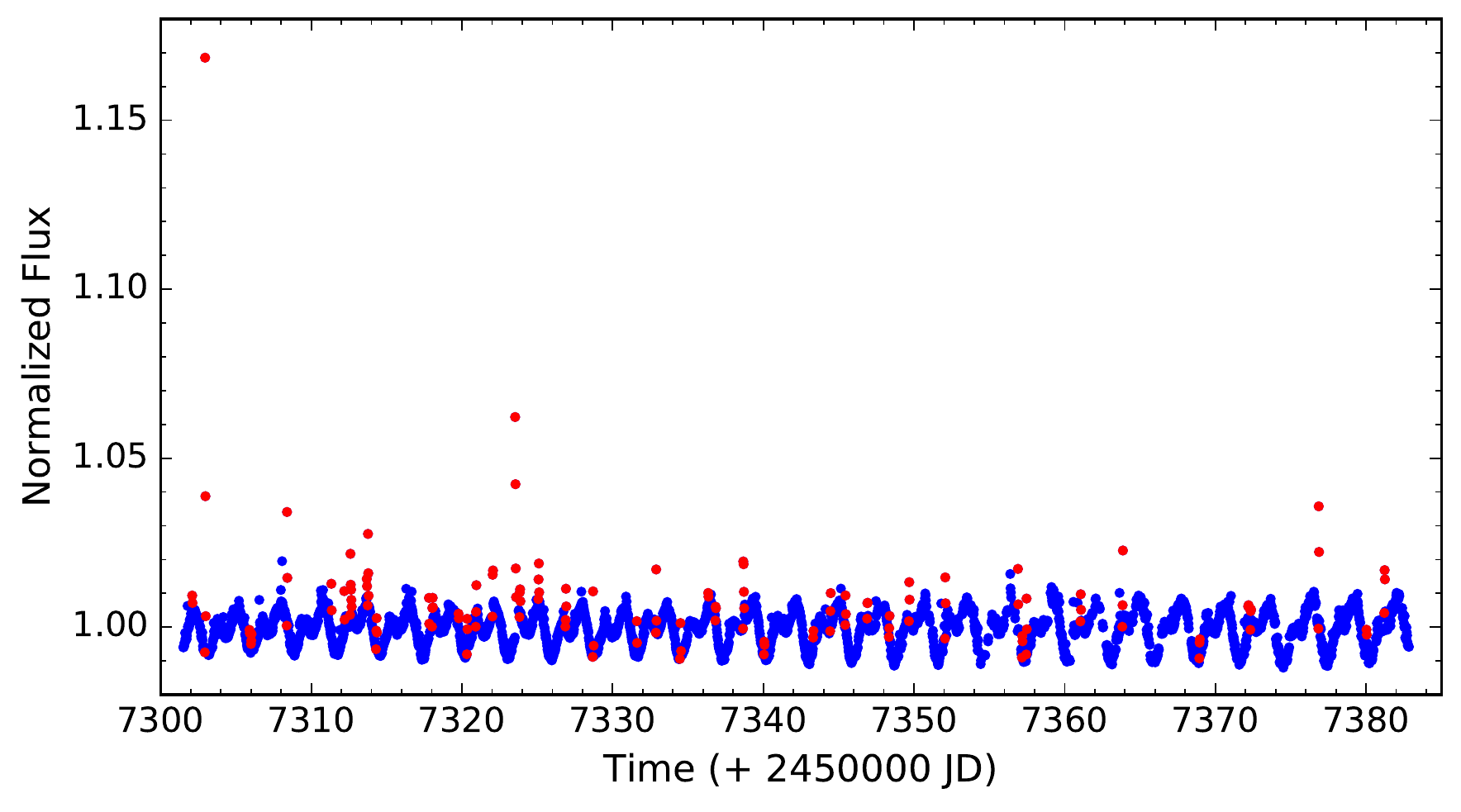}
       \caption{Long-cadence $K2$ photometric light curve for Gl 729. Selected flare candidates are shown with red points.}%
   \label{lc_rot_cf}
\end{figure*}

In Fig. \ref{lc_rot_cf} we show the photometric time series for Gl 729, where an appreciable rotational modulation can be noted. 
The red points indicate the flares detected by  FLATW'RM for N=2 and a detection limit of 3$\sigma$. 
These parameters were selected according to the statistical characteristics of flares of active M stars. According to \cite{Gunther19},  the equivalent duration of a flare in an active M0-M4 or M4.5-M10 star lies between  $\sim$1 and $\sim$120 minutes. 
Based on the 30-minute cadence of the \textit{Kepler} light curve in Fig. \ref{lc_rot_cf}, we therefore requested at least two consecutive points above the detection limit to consider a single flare lasting at least 30  minutes. 
Shorter flares could not be detected with this sampling. We found a total of 47 flare events in $\sim$ 81 days. 
The flare energy can be estimated by integrating the flare-normalized intensity during the event between the beginning and the end times detected by the algorithm (see \citealt{Vida18} for more details). 
This equivalent duration $\varepsilon_f$ is multiplied by the quiescent stellar luminosity ($F_{\star}$), to obtain the energy in the observed passband ($\xi_f$).
We estimated the quiescent luminosity by integrating a  flux calibrated X-SHOOTER spectrum of Gl 729, convolved  with  the  \textit{Kepler}  response  function, to  obtain  the  observed quiescent  luminosity in the \textit{Kepler} pass band, and we obtain  $F_{\star} = 1.3 \times 10^{30}$ erg s$^{-1}$.

Following the analysis of \cite{Gizis17}, we studied the cumulative flare frequency distribution $\nu$ (i.e., the number of flares with a given energy or greater divided by the total time of observation in the light curve). It can be expressed as

\begin{equation}
\log \nu = a + \beta \log \xi_f.
\label{ec.cumf}
\end{equation}

\begin{figure*}
\centering
   \includegraphics[width=9cm]{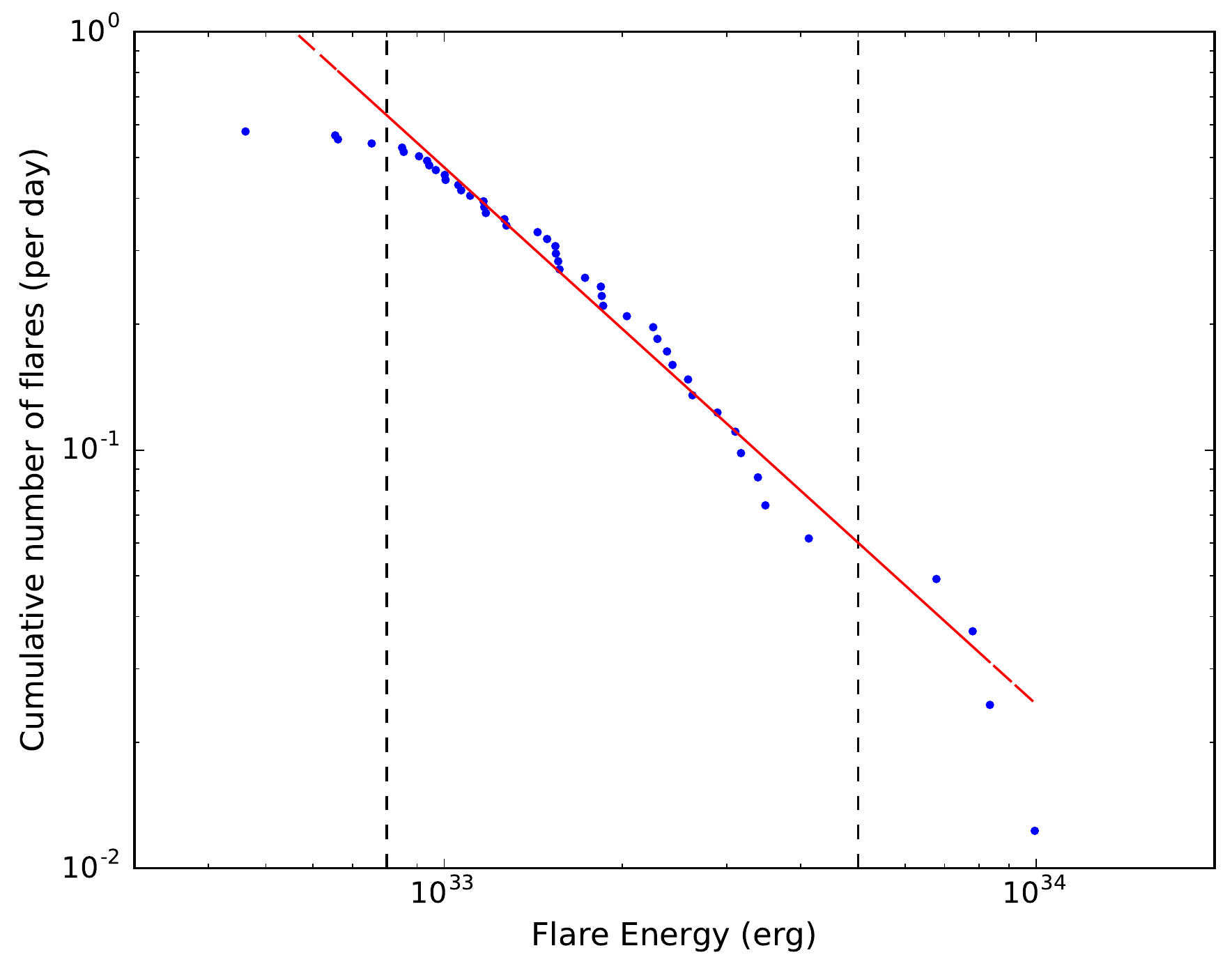}
       \caption{Gl 729. Cumulative frequency of flares $\nu$ derived from the \textit{Kepler} light curve with the FLATWR'M  algorithm. The red solid line is the best linear fit (Eq. \ref{ec.cumf}) in the range 8$\times$10$^{32}$-5$\times$10$^{33}$ erg with $\alpha=1.71$ %\noteRI{alpha Xshooter = 1.73, alpha fit = 1.71. Alpha de 1.7 parece que es calentamiento no térmico, VER BIEN. Líneas punteadas, }
       }%
   \label{FFD}
\end{figure*}

The slope $\beta = (1 - \alpha)$ is found by fitting the distribution by a linear function, where $\alpha$ is used to characterize how the flare energy of the star is dissipated. 
In Fig. \ref{FFD} we show the best linear fit for the energy range indicated with dashed black lines. We obtained a value of $\alpha = 1.71$ from this least minimum-squares fit.
\cite{Gizis17} proposed an alternative method for calculating $\alpha$ using a maximum likelihood estimator for the small sample size,

\begin{equation}
(\alpha - 1) = (n-2) \left[ \displaystyle\sum_{i = 1}^{n} \ln{\frac{\xi_i}{\xi_{min}}}\right] ^ {-1}
\label{ec.cumf2}
,\end{equation}
where $n$ is the number of transient events and $\xi_i$ and $\xi_{min}$ are the individual and lowest flare energies, respectively. 
With this method, we obtain $\alpha = 1.73,$ which is consistent with the linear-fit value.

Thus, Gl 729 flares follow a power-law slope $\beta \sim 0.71$ between  $10^{32}$ and $10^{34}$ erg.
Considering that the energy flux in the quiescent state is around $10^{30}$ erg s$^{-1}$, the energy release during a flare event is larger by 2 to 4 orders of magnitude than the time-integrated stellar luminosity. 

\begin{figure*}
\centering
   \includegraphics[width=\textwidth]{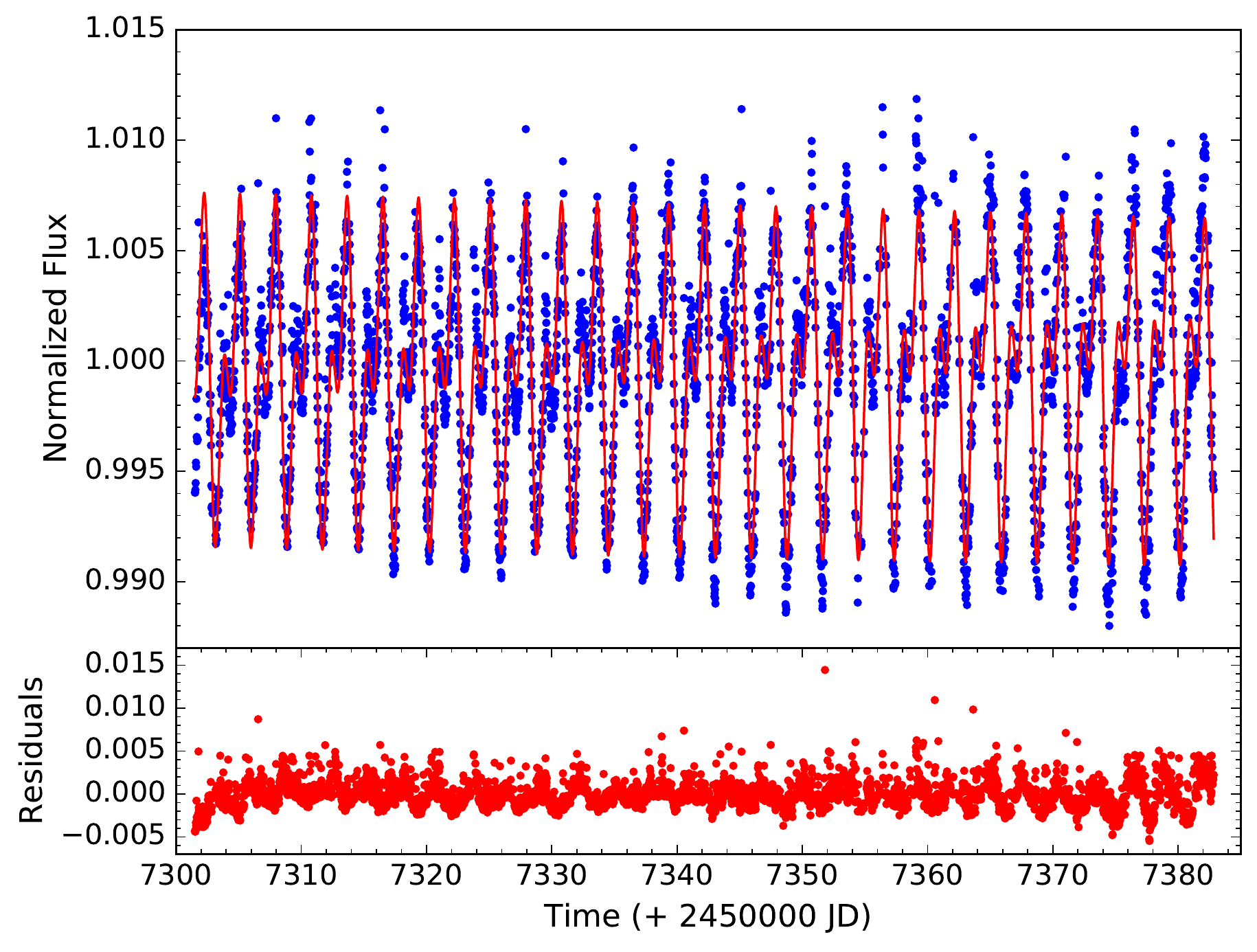}
       \caption{
        $K2$ long-cadence photometry for Gl 729 without the flare candidates detected with the FLATW'RM algorithm. The solid red line represents the least-squares fit with two harmonic function of the periods found with the GLS periodogram (2.848 days and 1.427 days).
       }%
   \label{lc_rot}
\end{figure*}

\begin{figure}
\centering
  \includegraphics[width=9cm]{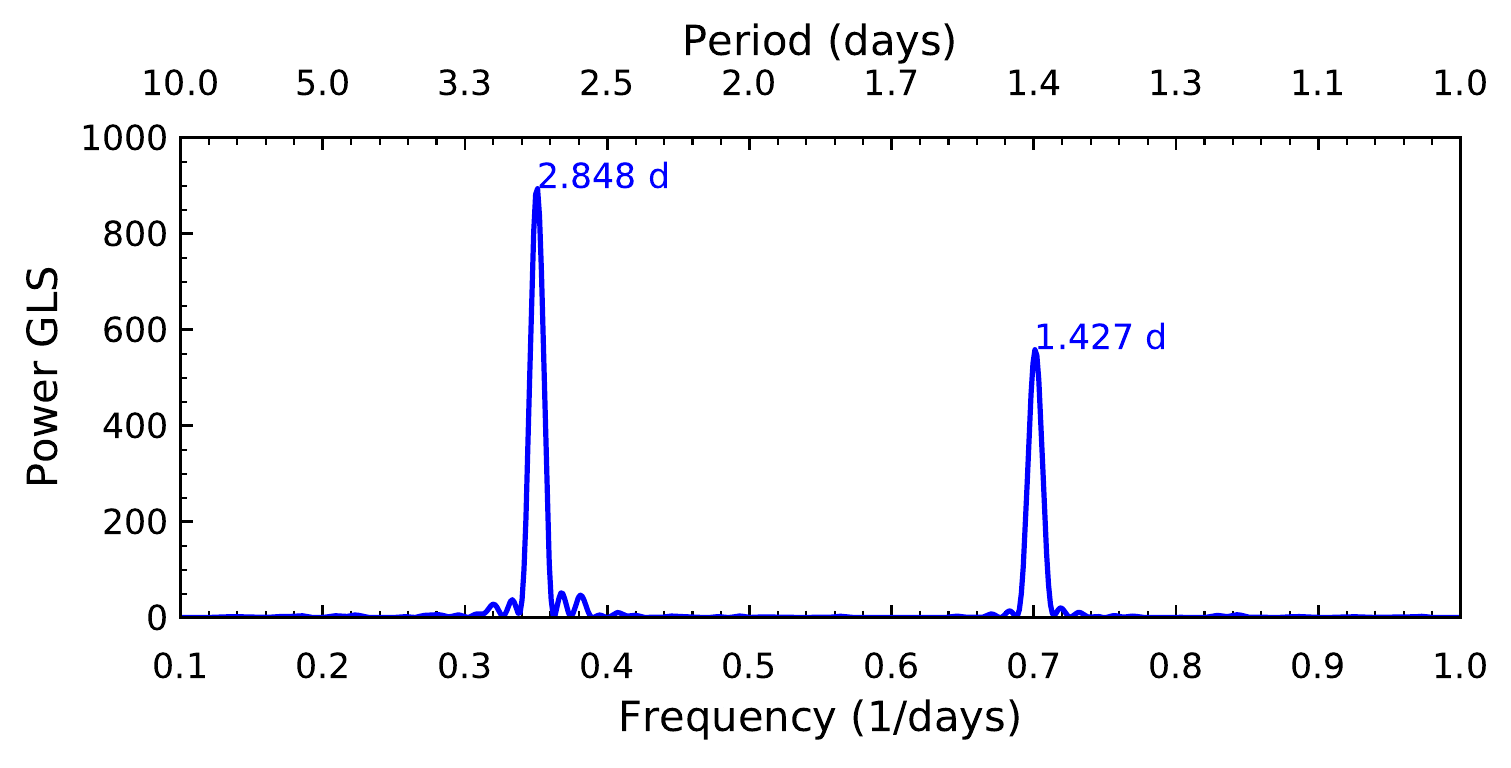}
   \caption{ GLS periodogram for the nonflaring $K2$ light curve of Fig. \ref{lc_rot}.
   }
   \label{per_gls-K2}
\end{figure}

To analyze the rotational modulation, we discarded the flares detected by FLATW'RM.  We studied the resulting light curve, shown in Fig. \ref{lc_rot}, using the generalized Lomb-Scargle (GLS) periodogram \citep{Zechmeister09}  in Fig. \ref{per_gls-K2}. 
The most significant period we detected is the rotation period found in the literature (see Section 2). We obtained $P_{rot} = (2.848 \pm 0.001)$ days with a high significance (false-alarm probability, FAP < 0.1 \%). We also found a second peak of $P = (1.427 \pm 0.001)$ days, half $P_{rot}$, with lower significance.
Following \cite{Lamm04}, we estimate the error of the detected period  as $\delta P=\frac{\delta\nu P^2}{2}$, where $\delta\nu$ is the finite frequency resolution of the periodogram. 
In Fig. \ref{lc_rot}  we also plot the best fit with two harmonic functions of these two periods with a red line. 

This bimodality could be associated to two dominant spots in opposite hemispheres, with areas $A_1$ and $A_2 < A_1$. This case would be equivalent to having a symmetric spot with area $A_2$ in both hemispheres that rotate with $P_{rot}/2$ and an asymmetric spot with area ${A_1-A_2}$ in only one hemisphere that rotates with $P_{rot}$. 
\cite{Mcquillan13} performed a statistical analysis of  \textit{Kepler} light curves of a series of M stars and found that  several stars  of their sample present this bimodality in their rotation periods.

\subsection{Spot modeling}
\label{methods}

A critical ingredient of the $\alpha\Omega$ dynamo is the differential rotation in the stellar interior (e.g., convection zone). A good proxy might be its surface differential rotation (see, e.g., \citealt{Buccino20}), which can be derived from the stellar spot  migration. 

When the light curve is phased with the average rotation period, the mean longitude of the activity centers at latitudes with rotation either shorter or longer than the average period are expected to migrate. 
Spot modeling is one possible approach to infer such a migration.

We applied a spot modeling approach that was introduced in \citet{Bonomo12}, to which we refer for details.
In brief, the surface of the star is subdivided into 200 surface elements that contain unperturbed photosphere, dark spots, and solar-like faculae. 
The specific intensity of the unperturbed photosphere in the \textit{Kepler} passband is assumed to vary according to a quadratic limb-darkening law,
\begin{equation}
I(\mu) = I_{0} (a_{\rm p} + b_{\rm p} \mu + c_{\rm p} \mu^{2}), 
\label{limbdark}
\end{equation}
where $I_{0}$ is the specific intensity at the center of the disk, $\mu = \cos \theta,$ with $\theta$ being the angle between the local surface normal and the line of sight, and $a_{\rm p}$, $b_{\rm p}$, and $c_{\rm p}$ are the limb-darkening coefficients in the \textit{Kepler} passband \citep{Claret11}. 

The dark spots are assumed to have a fixed contrast $c_{\rm s} \equiv I_{\rm spot}(\mu)/I(\mu)$ in the \textit{Kepler} passband, where $I_{\rm spot}$ is the specific intensity in the spotted photosphere. 
The fraction of a surface element covered by dark spots is given by its filling factor $f$. 

This model is fit to a segment of the light curve of duration $\Delta t_{\rm f}$ (see Sect.~\ref{parameters}) by varying the filling factors of the individual surface elements that can be represented as a 200-element vector $\vec f$. 
The spot pattern is assumed to stay fixed in each interval of duration $\Delta t_{\rm f}$, which is a fundamental assumption of our modeling because a significant spot evolution that would occur on a shorter timescale may hamper our approach.

Our model has 200 free parameters and is  nonunique and instable because of the effect of photometric noise. 
To select a unique and stable solution, we applied a maximum entropy regularization by minimizing a functional $Z$ that is a linear combination of the $\chi^{2}$ and of a suitable entropy function $E$, 
\begin{equation}
Z = \chi^{2}(\vec f) - \lambda E(\vec f),
\end{equation}
where $\lambda > 0$ is a Lagrangian multiplier that controls the relative weights given to the $\chi^{2}$ minimization and the configuration entropy of the surface map $E$ in the solution.  
The expression of $E$ is given in Eq.~(4) of \citet{Lanza98} and it is maximum  when the star is unspotted, that is, all the elements of the vector $\vec f$ are zero. 
In other words, the maximum entropy (hereafter ME) criterion selects the solution with the minimum spotted area compatible with a given $\chi^{2}$ value of the best fit to the light curve.   
When the Lagrangian multiplier $\lambda = 0$, we obtain the solution corresponding to the minimum $\chi^{2}$ that is unstable. 
By increasing $\lambda$, we obtain a unique and stable solution at the price of increasing the value of the $\chi^{2}$. 
An additional effect is that the residuals between the model and the light curve become biased toward negative values because we reduce the spot filling factors by introducing the entropy term (see \citet{Lanza16}, for more details).

The information on the latitude of the spots is lacking in our ME maps because the inclination of the stellar spin axis is very close to $90^{\circ}$ (cf. Sect.~\ref{parameters}), which makes the transit time of each  feature independent of its latitude. 
We therefore limit ourselves to map the distribution of the filling factor versus longitude.

The optimal value of the Lagrangian multiplier $\lambda$ is obtained by imposing that the mean $\mu_{\rm reg}$ of the residuals between the regularized model and the light curve verify the relationship \citep{Bonomo12,Lanza16}
\begin{equation}
| \mu_{\rm reg} | = \frac{\sigma_{0}}{\sqrt{N}},  
\end{equation}
where $\sigma_{0}$ is the standard deviation of the residuals of the unregularized model, that is, the model computed with $\lambda =0$, and $N$ the number of data points in the fitted light curve interval of duration $\Delta t_{\rm f}$.  

The optimal value of $\Delta t_{\rm f}$ is not known a priori and must be determined with an analysis of the light curve itself because it is related to the lifetimes of the active regions in a given star. 
We adopted a unique value of $\Delta t_{\rm f}$ for the entire light curve of Gl 729  because the ratio $\Delta t_{\rm f}/P_{\rm rot}$, where $P_{\rm rot}$ is the stellar rotation period, rules the sensitivity of the spot modeling to active regions located at different longitudes, as discussed by \citet{Lanza07}. 

The optimal value of the facular-to-spotted area ratio $Q$ could also be derived from the light curve best fit. 
Considering that at the young age of Gl 729  the activity is dominated by dark spots, and considering that  for any value of $Q$, we found highly structured maps that barely represent the double-dip shape of the light curve. We therefore decided to fix $Q$ = 0, that is, we included only dark spots in our model.

\subsection{Stellar parameters}
\label{parameters}

The basic stellar parameters, that is, mass, radius, and effective temperature $T_{\rm eff}$ , were taken from the references in Table\,\ref{model_param}. 
They do not directly enter our geometric spot model, except for the computation of the relative difference $\epsilon_{\rm rot}$ between the polar and equatorial axes of the ellipsoid used to represent the surface of the star.
Their values are obtained by a simple Roche model assuming rigid rotation with a period of $P_{\rm rot} = 2.848$ days. 
{The gravity-darkening effect associated with $\epsilon_{\rm rot} \sim  4.1 \times 10^{-5}$ is much smaller than the photometric precision}, thus it can be neglected in our model.
The inclination of the stellar rotation axis was assumed to be $i = 90^{\circ}$ because a star viewed equator-on is compatible with the stellar radius, the rotation period, and the projected rotational velocity $v.sin{i}$. 

The contrast of the dark spots c$_s$ = 0.90 was inferred from the work of \citet{Andersen15}. The duration of the individual segments of the light curves was kept at $\Delta t_{\rm f} = 2.848$ days, that is, the rotation period. 
We found that increasing $\Delta t_{\rm f}$  decreases the total $\chi^2$ . 
In contrast, keeping $\Delta t_{\rm f} = 2.848$ days, we obtain a better time resolution in the description of the spot evolution because the the timescale of the active region growth and decay in Gl 729 is comparable to or longer than the rotation period.  Choosing $\Delta t_{\rm f}$ equal to the rotation period is also optimal because it grants a uniform sampling of all the longitudes by our spot modeling \citep{Lanza07}. 

To infer information on the age, we checked possible membership with known associations or moving groups using the BANYAN $\Sigma$ tool \citep{Gagne18}. We found a 99.9\% probability that Gl 729 is a field star. 
Using relations between age, rotation period, and X-ray luminosity ($\log$ L$_X$ = 27.05; \citealt{Wright11}) calibrated for M2-M6 stars (\citealt{Engle18}; \citeyear{Engle11}), we infer an age in the range from 0.2\,Gyr to 0.6\,Gyr. This result is supported by the relatively high rotation velocity ($v \sin i = 3.5 \pm 0.5$ km s$^{-1}$) reported by \citealp{Johns96}), which indicates that Gl 729 is a young star with an estimated age younger  than 1 Gyr.

\begin{table}
\caption{Parameters adopted to model the light curves of Gl 729.}
\begin{center}
\begin{tabular}{lcc}
\hline
\hline
Parameter & Value & Ref. \\
\hline
Star mass ($M_{\odot}$)     & 0.14 & \citet{Gaidos14} \\
Star radius($R_{\odot}$)   & 0.19 & \citet{Gaidos14} \\
$T_{\rm eff}$ (K) & 3213  & \citet{Gaidos14} \\
$v\sin{i}$ (km s$^{-1}$)   &  3.5$\pm$0.5  & \citet{Johns96}\\
%$\log g $ (cm\,s$^{-2}$) & xxx & xxx \\
$a_{\rm p}$ & 0.2023 & \citet{Claret11} \\
$b_{\rm p}$ & 1.1507 & \citet{Claret11} \\
$c_{\rm p}$ & $-$0.3530 & \citet{Claret11} \\
$P_{\rm rot}$ (days) & 2.848 & present study\\
$\epsilon_{\rm rot}$ & $4.1\times 10^{-5}$ & present study \\
$i$ (deg) & 90.0 & present study \\
$c_{\rm s}$ & 0.90 & \citet{Andersen15}\\
%$c_{\rm f}$  & 0.115 & BL12 \\
$Q$ & 0.0 & present study \\
$\Delta t_{\rm f}$ (days) & 2.848 & present study\\
 \hline
\noindent
\end{tabular}
\end{center}
\label{model_param}
\end{table}

\subsection{Model results}

\begin{figure*}[htb!]
\centering
   \includegraphics[width=16cm, angle=0, trim = 0 500 0 0]{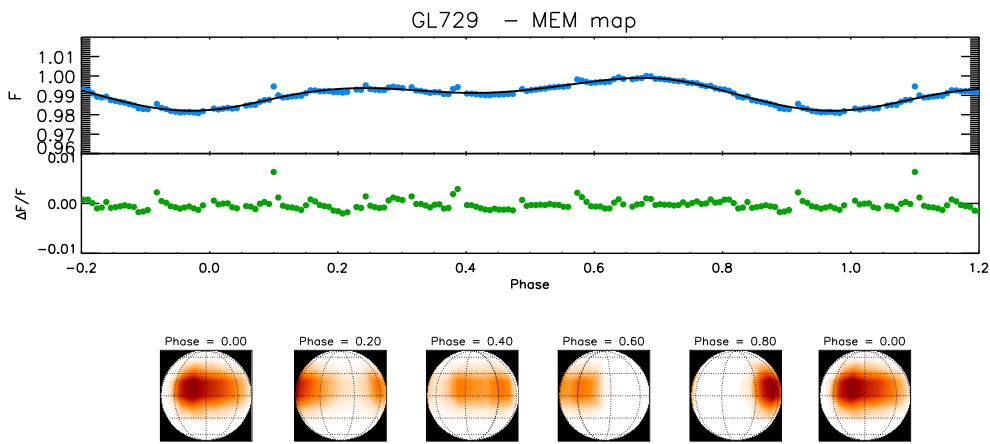}
       \caption{Example of spot modeling of a single stellar rotation.}%
   \label{best_fit_single}
\end{figure*}

As mentioned in Sect.\,\ref{methods}, we found the standard deviation $\sigma_0$ of the residuals of the unregularized model (obtained by imposing $\lambda = 0 $)  to be $\sim$ 1.1  $\times$ 10$^{-3}$ and the average number of data points in the fitted light curve interval of duration $\Delta$t$_f$ to be N $\simeq$ 110. 
After a number of trials, we found Eq.\,(4) to be satisfied with $|\mu_{reg}|$ $\sim$ 1.0  $\times$ 10$^{-4}$. 
In Fig.\,\ref{best_fit_single} we show an example of the results of our spot modeling for the epochs from BJD 2457340.26 until 2457343.11, spanning a single stellar rotation.

In the top panel we show the normalized flux with the best fit overplotted, in the middle panel the distribution of residuals, and finally, in the bottom panel the spot maps  at five selected rotation phases. 
Two major spot groups of different sizes in opposite hemispheres are clearly visible, which is compatible with the discussion in Sect.\,\ref{short-term}. 
We note that the \textit{Kepler K2} PDCSAP fluxes used in our analysis show evidence of residual instrumental effects, which are more clearly visible in the residual plot as discontinuities about every 0.1 intervals in rotation phase. However, the computed model (solid line) is apparently not affected by this issue.

\begin{figure*}[htb!]
\centering
   \includegraphics[width=12cm,height=16cm,angle=0]{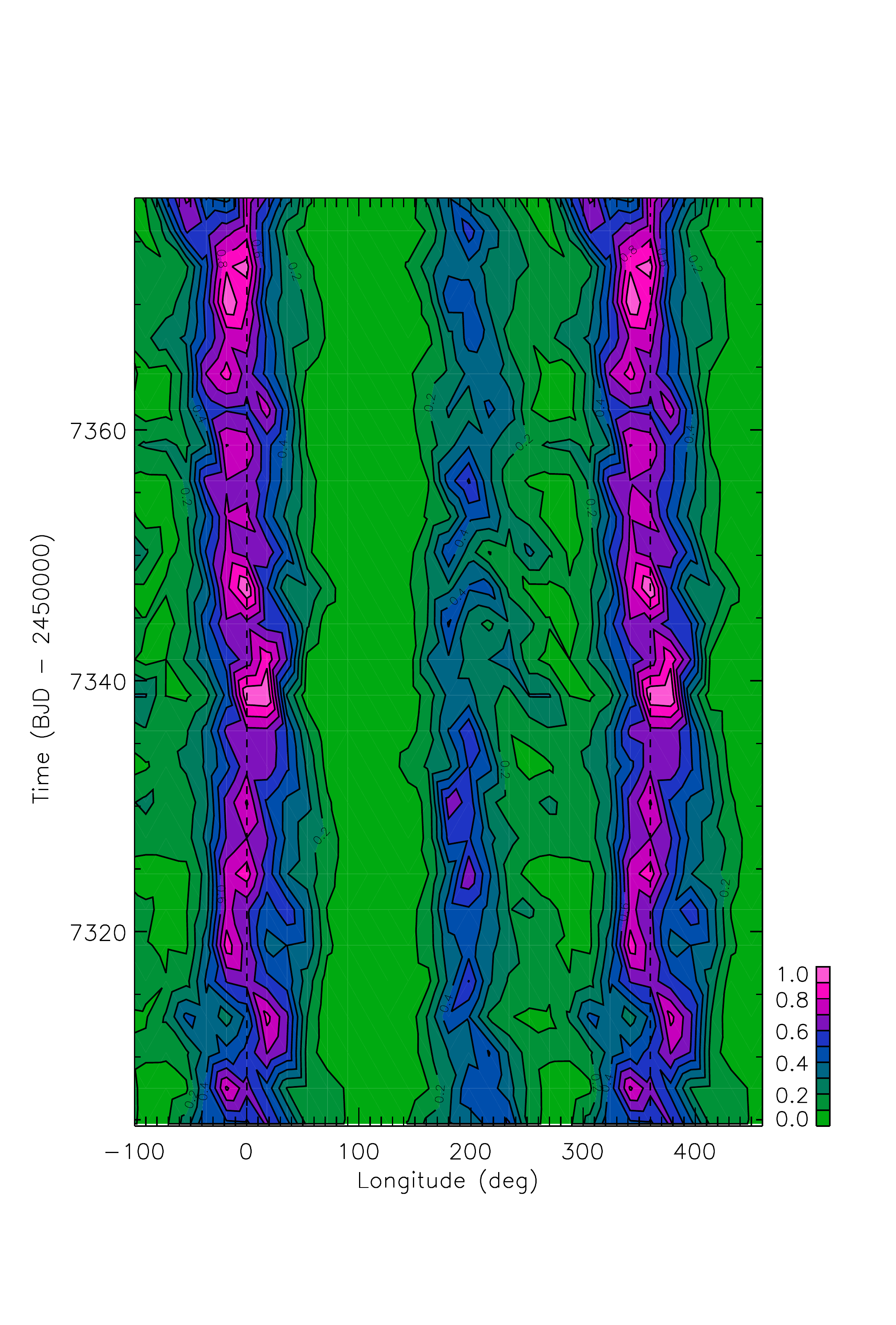}
       \caption{Distribution of the spot filling factor
vs. longitude and time as derived by
our maximum-entropy spot model. The maximum of
the filling factor is indicated in purple and the minimum in green
(see the color scale in the lower right corner).
We note that the longitude scale is repeated
beyond the $[0^{\circ}, 360^{\circ}]$ interval.}%
   \label{contour}
\end{figure*}

As anticipated, we are specifically interested in verifying the possible migration of the longitude at which spots are located. 
This is accomplished in Fig.\,\ref{contour}, where we plot the distribution of the filling factor of the starspots $f$ versus the longitude and the time for the regularized spot maps. 
The origin of the longitude is at the meridian pointing toward the Earth on BJD 2457303.243, and the longitude increases in the same direction as the stellar rotation. 
We identify two activity centers, a dominant one located at about longitude zero, and another smaller center in the opposite hemisphere. 
These centers could be associated to the two peaks observed in the GLS periodogram of Fig. \ref{lc_rot}, one corresponding to the rotation period and the other to its first harmonic.
We note some indication of an oscillation in the longitude of the dominant activity center. 
However, the amplitude of this migration, which is about 50$^{\circ}$ , is comparable to or smaller than the longitude resolution achieved by our spot modeling. 

Our spot modeling allows us to determine the variation in total spotted area versus time by integrating the filling factor over the longitude.  
The error is estimated from the photometric accuracy of the data points. 
The gaps inside each individually fit interval of duration $\Delta t_{\rm f}$ affects the total area because the maximum entropy regularization drives the solution toward the minimum spotted area compatible with the data, thus reducing the filling factor at the longitudes that are in view during the gaps in the light curves. 

To reduce this effect on the variation of the entire spotted area, we measured the presence of significant gaps in each interval $\Delta t_{\rm f}$. 
We divided each interval into five equal subintervals and counted the number of data points in each subinterval $n_{i}$, with $i=1,..,5$ numbering the subinterval. 
A measure $\delta$ of the inhomogeneous distribution of the data points in the interval $\Delta t_{\rm f}$ is defined as $\delta \equiv [\max (n_{i})-\min (n_{i})]/ \max (n_{i})$. 
The intervals with $\delta > 0.2$ are discarded, giving a total of 21 area measurements that are not affected by the gaps in a total of 27 intervals.

\begin{figure}[htb!]
\centering
   \includegraphics[width=9cm, angle=0]{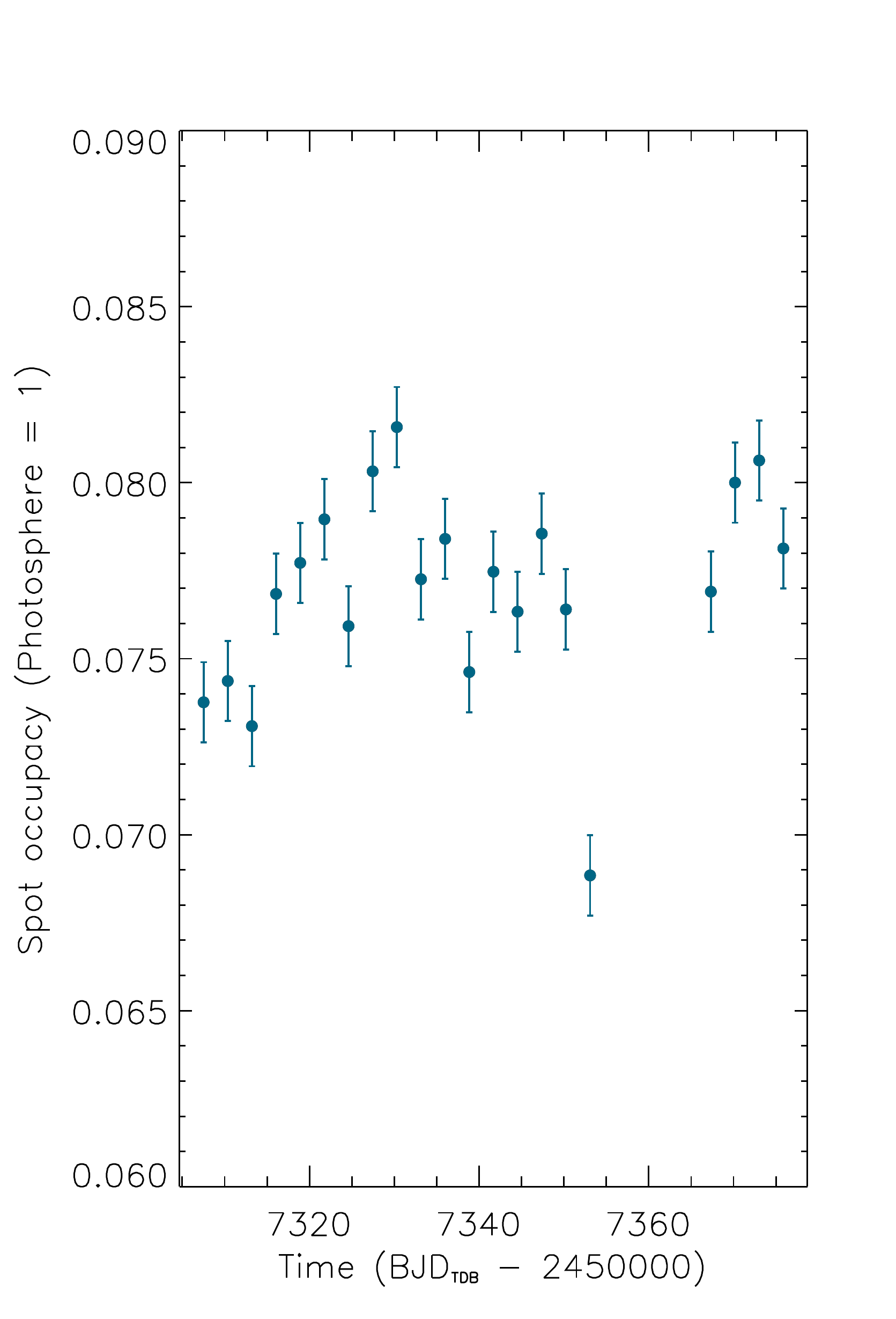}
       \caption{Total spotted area as derived from the ME best fits to the light curve vs. time.  The error bars have an amplitude of $3\sigma$, where $\sigma$ is the standard deviation as derived from the photometric accuracy of the data points.  }%
   \label{spot_area}
\end{figure}

The plot of the total spotted area versus the time for this light curve is shown in Fig.~\ref{spot_area}.  
The duration of the time interval is too short to conclude about the cause of the area variation, which might be associated with the growth and decay of the individual active regions in the active longitudes.

%__________________________________________________________________
%__________________________________________________________________
%__________________________________________________________________

\section{Chromospheric activity}\label{sec.spec}

In this section we analyze the long-term activity of the flare active M4 dwarf, Gl 729. We present a magnetic activity analysis by compiling our own and public spectroscopic data and building a registry of activity that allows us to detect cyclic patterns.

\subsection{Mount Wilson $S$-index}

Stellar activity cycles have been detected in several late-type stars, typically by measuring fluctuations in the well-known dimensionless  Mount Wilson $S$-index (e.g., \citealt{Baliunas95,Metcalfe13,Ibanez18}). 
This activity indicator $S$  is defined as the ratio between the chromospheric Ca \scriptsize{II}\normalsize\- H and K line-core emissions, integrated with a triangular profile of 1.09 \AA\- full width at half maximum (FWHM), and the photospheric continuum fluxes integrated in two 20 \AA\- passband centered at 3891 and 4001 \AA\-  \citep{Duncan91}. 
For decades, the $S$-index was mainly used to study the chromospheric activity only for dF to dK stars \citep{Baliunas95} because longer exposure times are needed to observe the Ca \scriptsize{II}\normalsize\- lines in later stars, which are both redder and fainter. \cite{Cincunegui07b} studied the usefulness of the S index and other activity indicators for dM stars.

\begin{figure*}
\centering
   \includegraphics{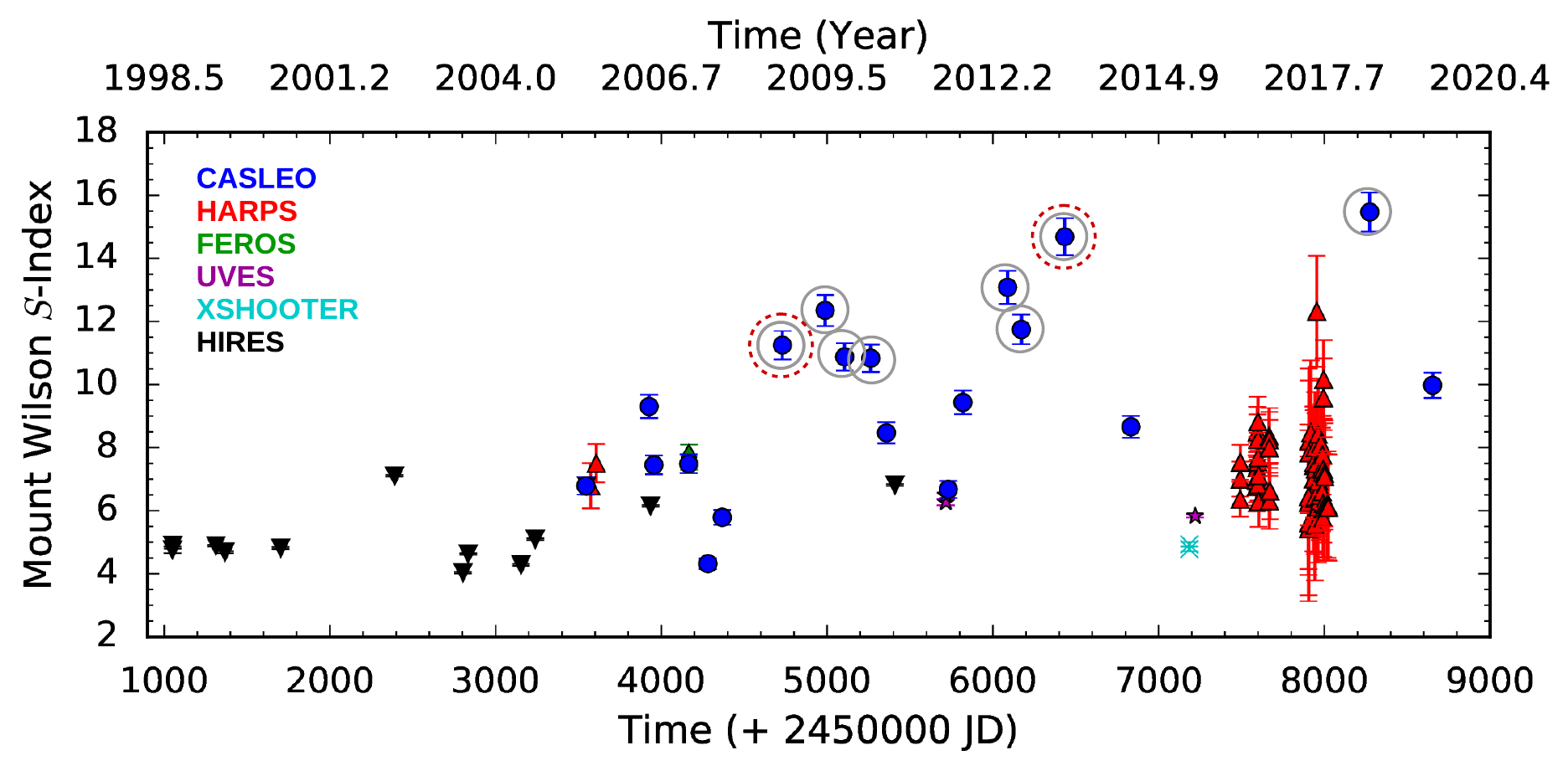}
       \caption{Mount Wilson $S$-indexes for Gl 729 derived from CASLEO (blue circles), HARPS (red triangles), FEROS (green diamonds), UVES (magenta stars), XSHOOTER (cyan cross), and HIRES (black inverted triangles) spectra. We highlight the CASLEO spectra where an increase in activity is observed with gray circles. We show the two observations we discarded from our analysis as dashed red circles.}%
   \label{st_CHFUXI}
\end{figure*}

To search for indications  of stellar activity in Gl 729, we computed the $S$-index for the spectra mentioned in Sect. \ref{sec.obs}. For our CASLEO observations, we used the method described in \cite{Cincunegui07b}. For the other spectra, we followed \cite{Duncan91}. 
We then calibrated the HARPS indexes to the Mount Wilson $S$-index with the calibration available in \cite{Lovis11}, and the FEROS spectra with the procedure used in \cite{Jenkins08}. 
Finally, as we have done in previous work \citep{Ibanez18,Ibanez19}, we intercalibrated the CASLEO, FEROS, UVES, XSHOOTER, and HIRES indexes considering as reference the calibrated HARPS Mount Wilson indexes closest in time.
Our time series is composed of 99 measurements for a time span between 1998 an 2019. These data are shown in Fig. \ref{st_CHFUXI} and listed in Table \ref{obs_table}. 

We estimated a 4\% typical error of the $S$-index  derived  from CASLEO (see the deduction in \citealt{Ibanez18}). 
The error bars of the $S$-indexes derived from  HARPS, FEROS, UVES, XSHOOTER, and HIRES spectra were calculated as the  standard deviation of each monthly bin. 
The rotational modulation in the $S$ -index may contribute to its daily variation; we estimated it to be $\sim15\%$ from HARPS observations.
For time intervals with only one ESO observation in a month, we adopted the typical RMS dispersion of the  bins. 
The whole time series presents a variability of $\sigma_{S}$/$\langle S \rangle$ $\sim$ 27 \% in 21 years.

\subsection{Long-term variations}

The time series plotted in  Fig. \ref{st_CHFUXI} shows an almost flat regime  between1998 and 2004, and then
an increasing trend until the maximum in 2018 (xJD = 8300 days). 
First, we analyzed if this  growth is due to the gradual increase of the mean magnetic activity or to particular observations of transient high-energy phenomena (e.g., flares). Gl 729 has a flare frequency of  0.5 flares per day of at least 10$^{33}$ erg (see Fig. \ref{FFD}). To filter these events out, we visually compared the individual spectra that form each of our observations (see Section \S\ref{sec.obs})  
for the dates marked with gray circles in Fig. \ref{st_CHFUXI}.

\begin{figure}
\centering
   \includegraphics[width=9cm]{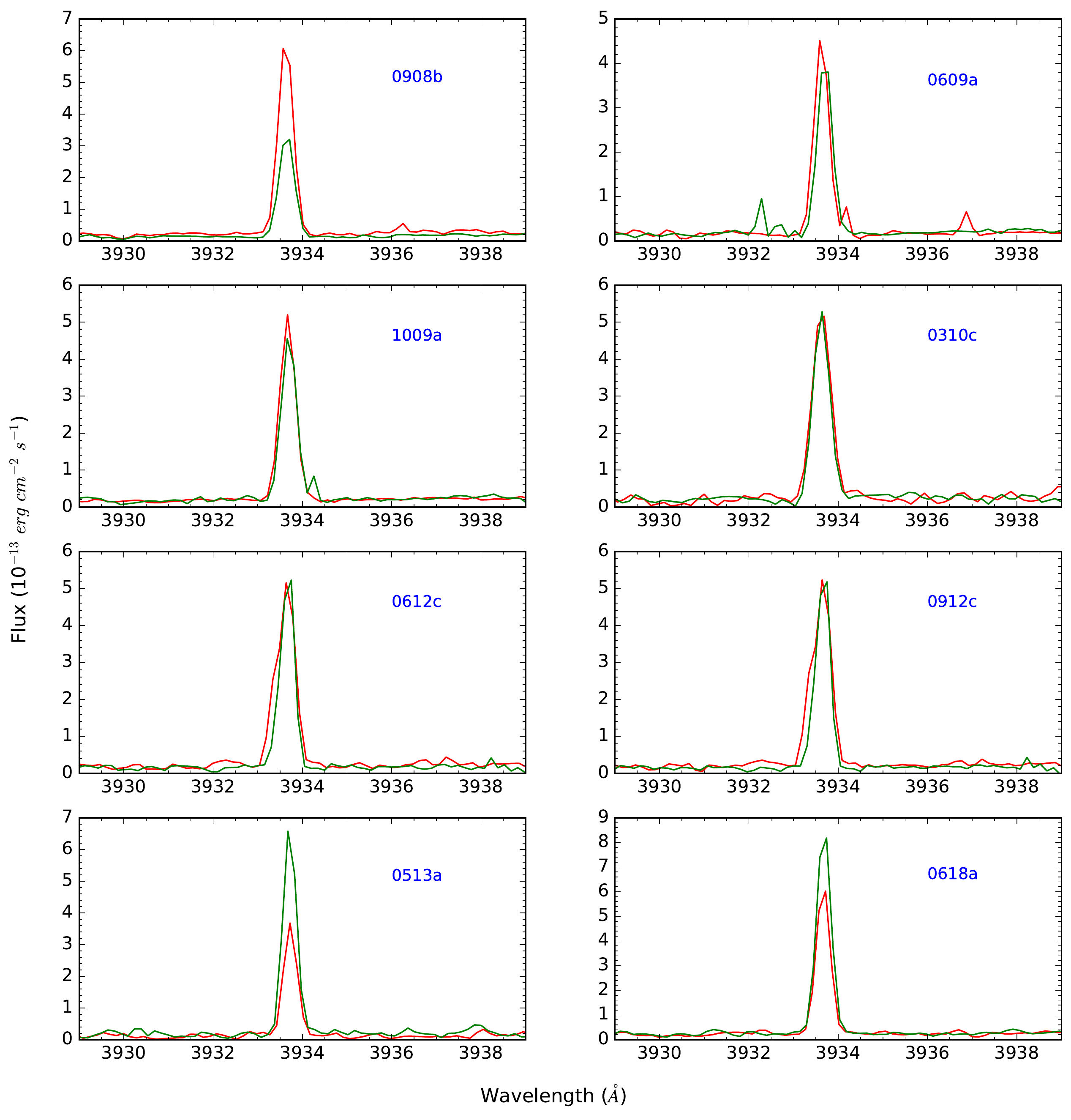}

       \caption{Ca \scriptsize{II}\normalsize\- K line for individual CASLEO observations. Each plot is labeled with the date of each observation (MMYY). We plot the first observation in red and  the second consecutive observation in the same night in green. 
       }%
   \label{indiv}
\end{figure}

In Fig. \ref{indiv} we show the individual line profiles of the Ca \scriptsize{II} \normalsize\- K line for these observations. We show in red the first and in green the second observation. 
There is a difference of 85\% between the two spectra taken in September 2008 (0908), and a difference of 78\% between those taken in May 2013 (0513). We excluded these two observations from our analysis because they were probably obtained during flares. For the observation taken in June 2018 (0618) we see a difference of almost 38\% between individual spectra. In this case we decided  to include only the first observation (red line in Fig. \ref{indiv}) because its line flux is similar to that of the remaining nonflaring observations. 
The difference between the two fluxes in the other observations ranges from  6\%\ to 11\%, consistent with the calibration error \citep{Cincu04}. 

\begin{figure}
\centering
   \includegraphics[width=9cm]{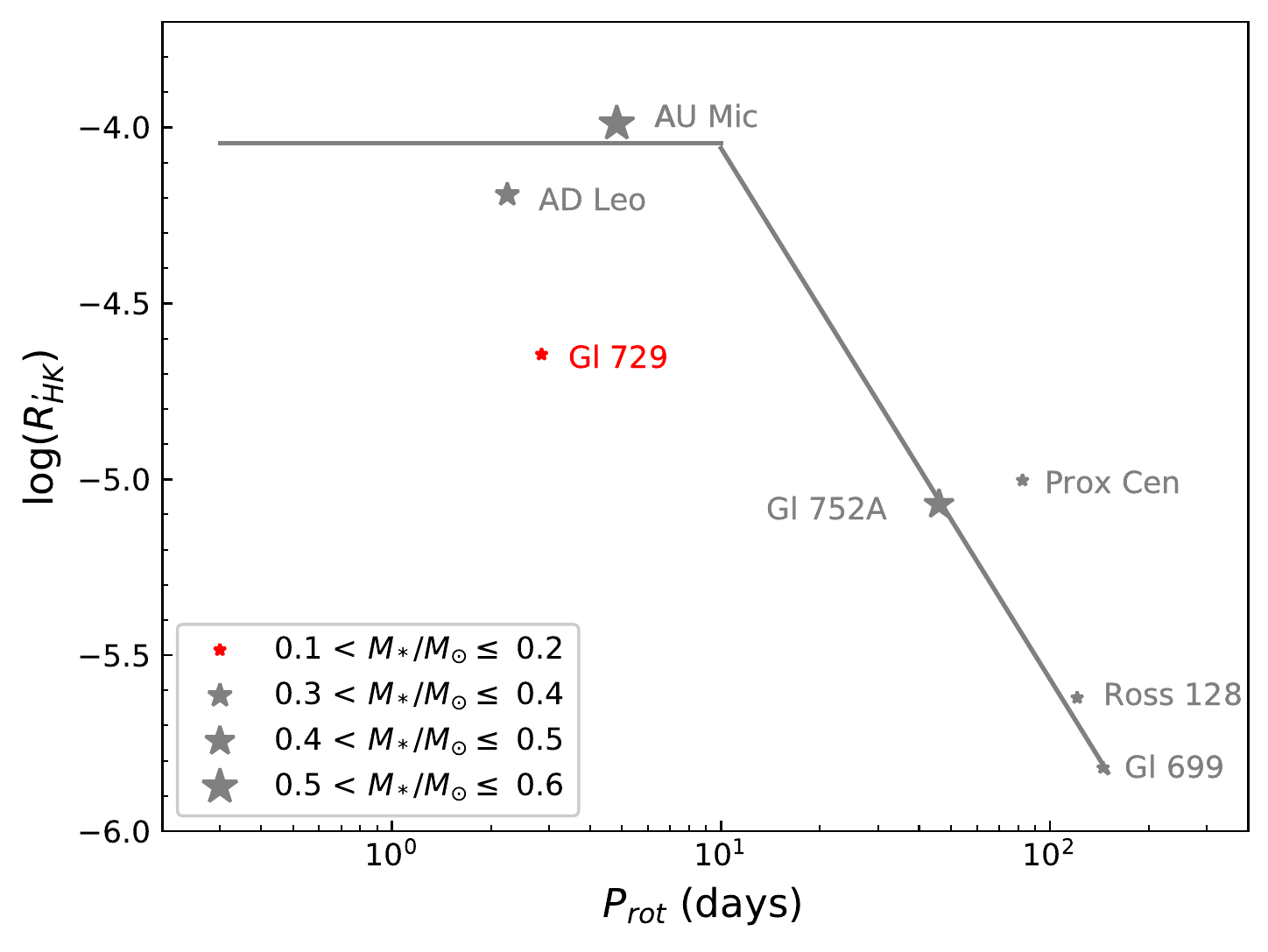}

       \caption{ $\log R'_{HK}-P_{rot}$ diagram including the fit of \citeauthor{Astudillo17} (gray solid lines). We show five M dwarfs whose activity cycles were detected using CASLEO spectra with gray stars (Prox Cen: \cite{Cincunegui07}; Gl 752A: \cite{Buccino11}; AD Leo: \cite{Buccino14}; AU Mic: \cite{Ibanez18}, and Ross 128: \cite{Ibanez19}). The activity cycle of Gl 699 was detected by \cite{Toledo19} employing seven independent sets of spectroscopic observations. We represent Gl 729 with a red star. It is a clear outlier from the saturated regime of this diagram.
       }%
   \label{a-d_diagram}
\end{figure}

From the remaining data, we obtained a mean Mount Wilson index $\langle S \rangle = 7.343 \pm 1.967$ and a Ca \scriptsize{II}\normalsize\- emission level of $\log R'_{HK} = -4.645$, which both agree with the values reported by \cite{Astudillo17}. 
Considering the rotation period for Gl 729 of 2.848 days, the  $\log R'_{HK}$ confirms that this star is an outlier in the saturation regime reported for dM stars in the $\log R'_{HK}-P_{rot}$ diagram  \citep{Astudillo17}. 
In  Fig. \ref{a-d_diagram} we plot the fit obtained by \cite{Astudillo17} from the HARPS database. We also include five stars for which magnetic activity cycles were detected employing CASLEO spectra and  Gl 699 (Barnard’s star), whose activity cycle was reported by Toledo-Padrón et al. (2019) using a different set of spectroscopic data.

\begin{figure}
\centering
  \includegraphics[width=9cm]{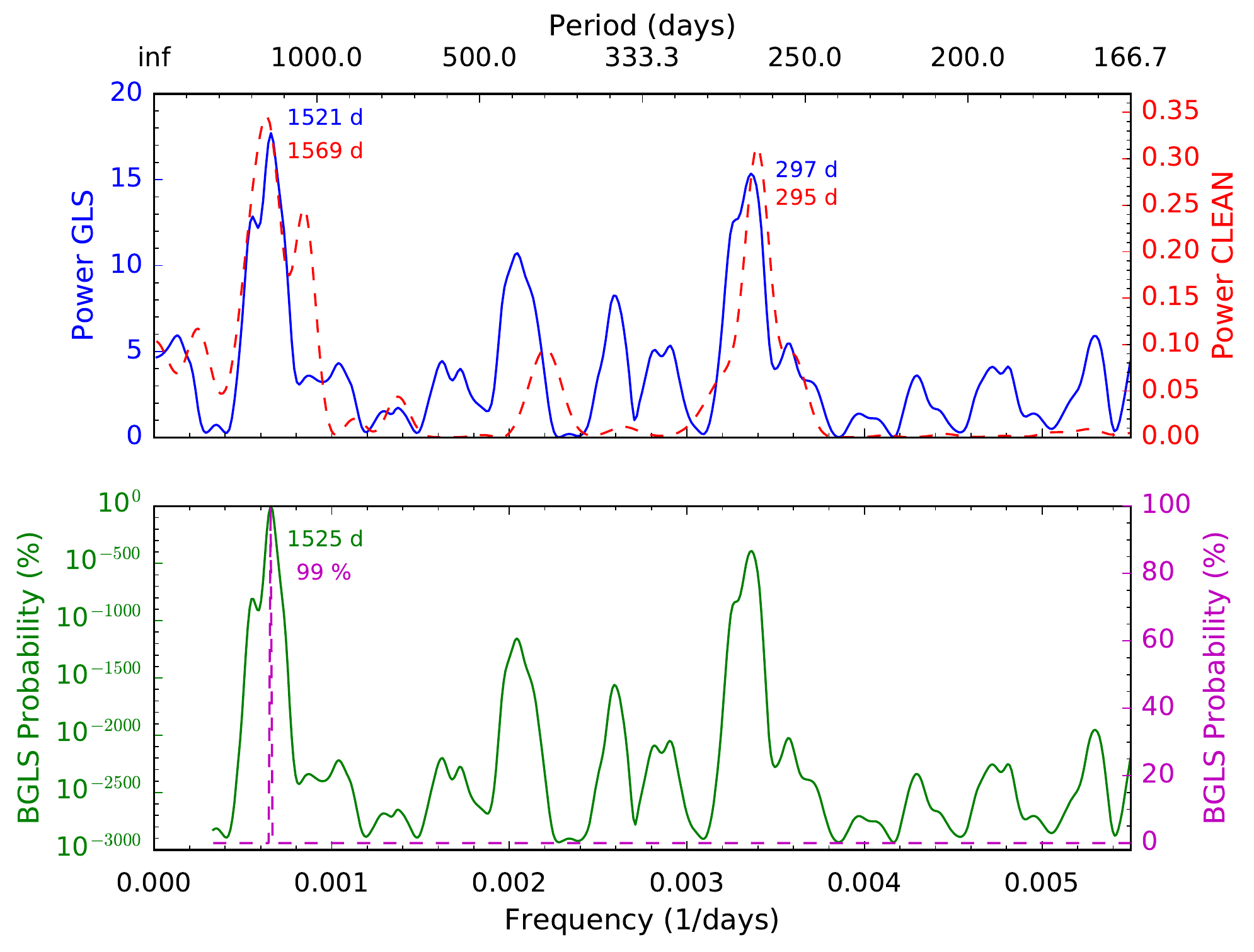}
   \caption{ \textit{Top:} GLS (solid blue line) and CLEAN (dashed red line) periodograms for the $S$ time series of Gl 729. 
   The most prominent peaks for the GLS periodogram with an FAP < 0.1\% are $(1521 \pm 20)$ days and $(297 \pm 2)$ days.
   \textit{Bottom:} Bayesian percentage probability indicates the 99 \% probability that the $\text{approximately}$ 1500-day peak is the real activity cycle.
   }
   \label{per_gls-bgls}
\end{figure}

To search for long-term activity cycles in the remaining data series, we implemented three different methods. In the top panel of  Fig.\ref{per_gls-bgls}
we show the GLS periodogram together with the results of the CLEAN deconvolution algorithm developed by \cite{Roberts87}.
Both analyses show a very significant peak at $P_1 = (1521 \pm 20)$ days for the GLS, a slightly higher value for the CLEAN periodogram, and a less significant one at $P_2 = (297 \pm 2)$ days, both with  an FAP < 0.1 \%.

In Fig. \ref{per_gls-bgls} (bottom) we present the results of the Bayesian generalized Lomb-Scargle periodogram described by \cite{Mortier15}, which expresses ``the probability that a signal with a specific period is present in the data''. 
The green line indicates the logarithmic probability, and the dashed magenta line shows the linear probability. 
We therefore conclude that both cycles are present in the data, although the $\text{about }$1500-day activity cycle, with a 99\% probability, is markedly more significant than the cycle with $ P = 296$ days.

\begin{figure}
\centering
  \includegraphics[width=9cm]{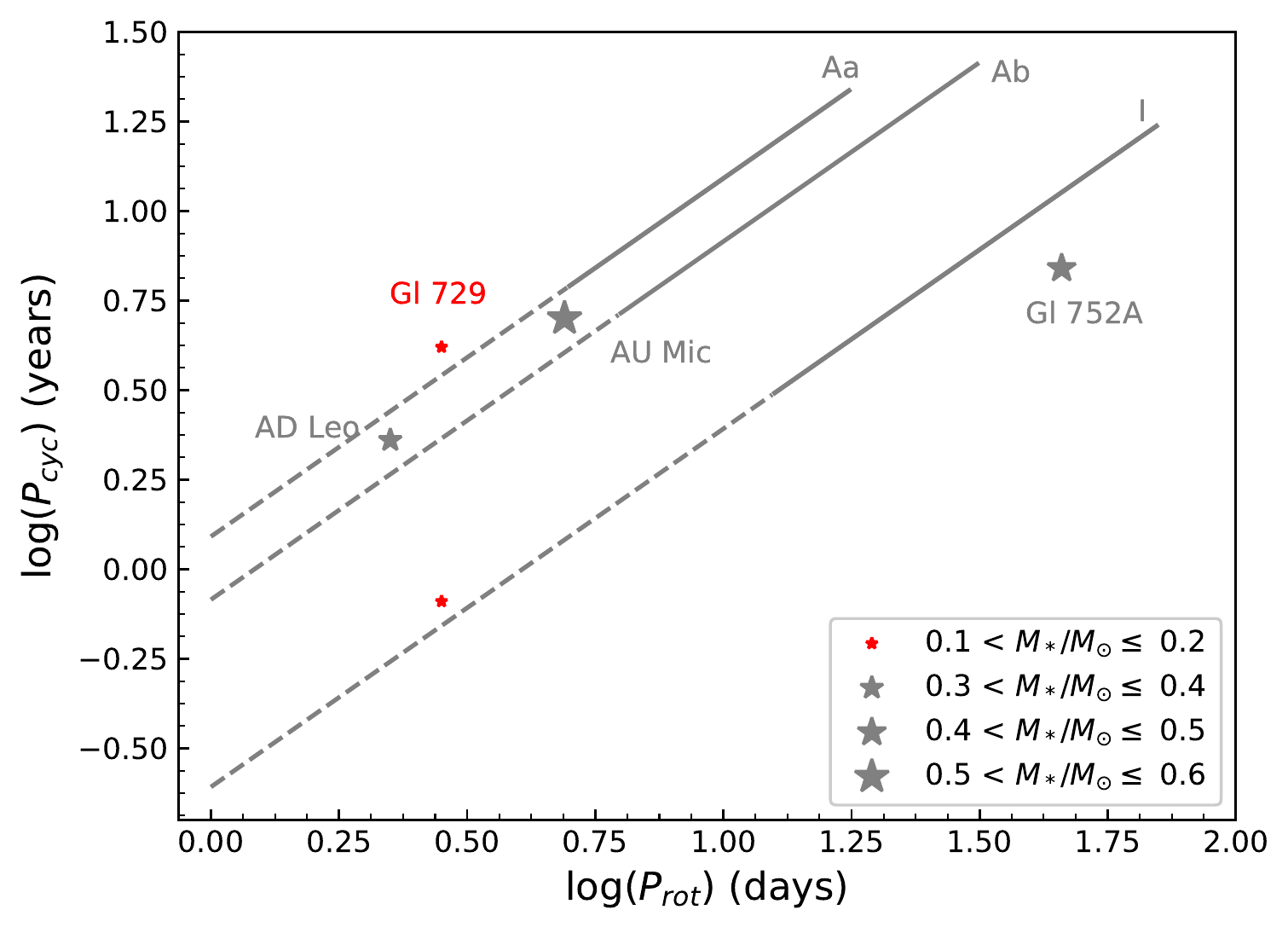}
   \caption{$\log P_{cyc}$ vs. $\log P_{rot}$. The solid gray lines represent the actives (Aa and Ab) and inactive (I) branches reported in \cite{Bohm07}. We indicate the extrapolation to lower rotation periods with dashed lines. Gray stars represent the activity cycle periods obtained for the early-M dwarfs Gl 752A \citep{Buccino11}, AD Leo \citep{Buccino14}, and AU Mic \citep{Ibanez18}. The red stars represent the periods of Gl 729 we report here.
   }
   \label{b-v_diagram}
\end{figure}

\cite{Bohm07} examined the relation between $P_{rot}$ and $P_{cyc}$ for a set of cyclic FGK stars, expanding the work done previously  \citep{Brandenburg98,Saar99}. She found that most stars are well distributed in two branches in the $P_{cyc}$ -- $P_{rot}$ diagram (her Fig 2). A possible interpretation is that each sequence corresponds to a different type of dynamo.
When we include our results in this $P_{cyc}$ -- $P_{rot}$ diagram, we find that the 1500-day cycle belongs to the active branch labeled ``Aa''  in that figure. For the stars in this  branch, the number of  rotational revolutions per activity cycle is about $P_{cyc}/P_{rot} \sim 500$, which is consistent with the value of $1521/2.848 = 534$ we obtain for Gl 729. 
Similarly, the less significant activity cycle of $P_{cyc}\sim 300$ days lies within the inactive branch, where the ratio $P_{cyc} / P_{rot} \sim 90$. Furthermore, as we have shown in \S\ref{parameters}, Gl 729 is a young star, which is coherent with the conclusion by  \cite{Bohm07} that  stars that belong to the Aa sequence are younger than the Sun. 
This suggests that \citeauthor{Bohm07}'s diagram can be extended to cyclic M stars. 
In Fig. \ref{b-v_diagram} we show the updated version of Fig. 2 from \cite{Bohm07}, and we include one dM and three dMe stars for which we found magnetic activity cycles.

\subsection{Wavelet analysis}

To explore the strength of the periodic signals we found over time, we performed a wavelet analysis  of the seasonal mean $S$-index measurements by implementing the wavelet analysis described in \cite{Torrence98} with the correction developed in \cite{Liu07}.
It consists of repeatedly convolving a selected waveform,  commonly called the mother wavelet,  with the data at each time step, using a range of scales for the waveform. 
Each scale is associated with a different frequency. 
The power obtained in each convolution produces a frequency map against the time domain.

We used the Morlet wavelet, which is a sinusoidal signal with a Gaussian amplitude modulation. 
By sliding it along the time series and changing the scale of the wavelet (i.e., its frequency or period), we obtained the wavelet power spectrum (WPS) as the result of the correlation between the wavelet and the data.

\begin{figure}
\centering
   \includegraphics[width=9cm]{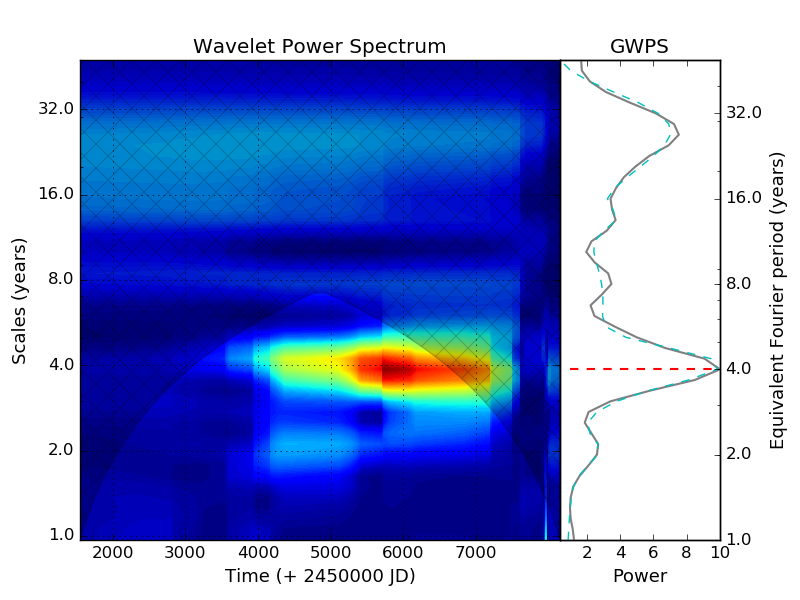}
       \caption{Contour wavelet plot for Gl 729, calculated using the $S$ time series of Fig. \ref{st_CHFUXI}. The GWPS is plotted to the right and is fit by a sum of Gaussians (dashed cyan line). The horizontal dashed red line in the GWPS represents the maximum peak.}%
   \label{wave}
\end{figure}

Fig. \ref{wave} shows the WPS for the time series of Fig. \ref{st_CHFUXI}.
The region delimited by the cross-hatched area is the wavelet region in which the edge effects become important, generally defined as  the cone of influence. 
Inside the cone, the power spectrum is shown with a color scale going from the weakest (dark blue) to the strongest (red and brown) signals. 
We note a  significant periodic signal between three and five years.

We also obtained the global wavelet power spectrum (GWPS), defined as
the sum of the WPS over time for each period of the wavelet. We show the GWPS at the right side of the Fig. \ref{wave}.
Following \cite{Garcia14}, we fit the GWPS with a sum of Gaussian curves associated with each peak.
The extracted global period corresponds to the highest amplitude peak, and its uncertainty is the half-width at half maximum (HWHM) of the corresponding Gaussian profile.
The horizontal dashed red line in the GWPS represents the most significant peak $P_{wave} = 4.1 \pm 0.7$ yr ($1497 \pm 256$ days),  in agreement with the 1521-day period detected with the periodograms above.

Additionally, the wavelet analysis allows us to study the distribution of the cyclic signal throughout the time series. 
Fig. \ref{wave} shows that the four-year period maintains its strength for most of the observations. 
However, before xJD $\sim$4000  days, the magnetic activity remains almost constant and the periodic signal is nearly insignificant (see Fig. \ref{wave}). 
At about  xJD=4000 days, the activity level of Gl 729 increases and the cyclic activity becomes more evident, until it reaches a peak after xJD $\sim$ 6000,  observed in red in the WPS map in Fig. \ref{wave}. Therefore we see here the same behavior as in Fig. \ref{st_CHFUXI}.

This phase of flat activity in the range xJD=1000-4000 days  in Gl 729 resembles the well-known  solar Maunder or Dalton Minima, although
the sampling in this interval is rather poor and  we are probably missing short term variations. 
Similarly, other solar-type stars also present this behavior: a broad activity minimum has also been observed in the K2V star $\varepsilon$ Eridani \citep{Metcalfe13} and in the G2V star HD 140538 \citep{Radick18}. However, this is the first time that it is reported for an M star.
Another possible interpretation is that a decadal activity cycle modulates the four-year cycle, although further observations of Gl 729 are required to  confirm this hypothesis.

\subsection{Sodium and $H{\alpha}$ as activity indicators.}

The Ca \II\- lines are not the most adequate feature to study chromospheric activity in M stars, which are too faint in this region of the spectrum.
To overcome this problem, it is necessary to  explore redder activity indicators in these stars. 
For instance, the  H$\alpha$ line has been extensively used as an activity indicator  \citep{Giampapa86, Cincunegui07b,Robertson14b}, as were other lines, for example, the Na \scriptsize{I}\normalsize\- D lines (\citealt{Diaz07b, GomesdaSilva12}). 

The analysis of different chromospheric lines does not only allow us to study magnetic activity at different heights of the atmosphere, but also to understand the energy transport in these regions for different activity levels in M stars.
In particular, atmospheric models of M stars show  that the Ca \scriptsize{II}\normalsize\- K line is formed in the lower chromosphere, while the H${\alpha}$ line is formed in the upper chromosphere with a different formation regime  \citep[e.g.,][]{ Mauas94,Mauas96, Mauas97, Fontenla16}.
\cite{Walkowicz09} observed a strong positive  correlation between single observations of the Ca \II\- lines and H$\alpha$ in most M3 V active stars.
However, this relation is not always valid for multiple observations of an individual active star \citep{Buccino14}. 

\begin{figure}
   \centering
      \includegraphics[width=9cm]{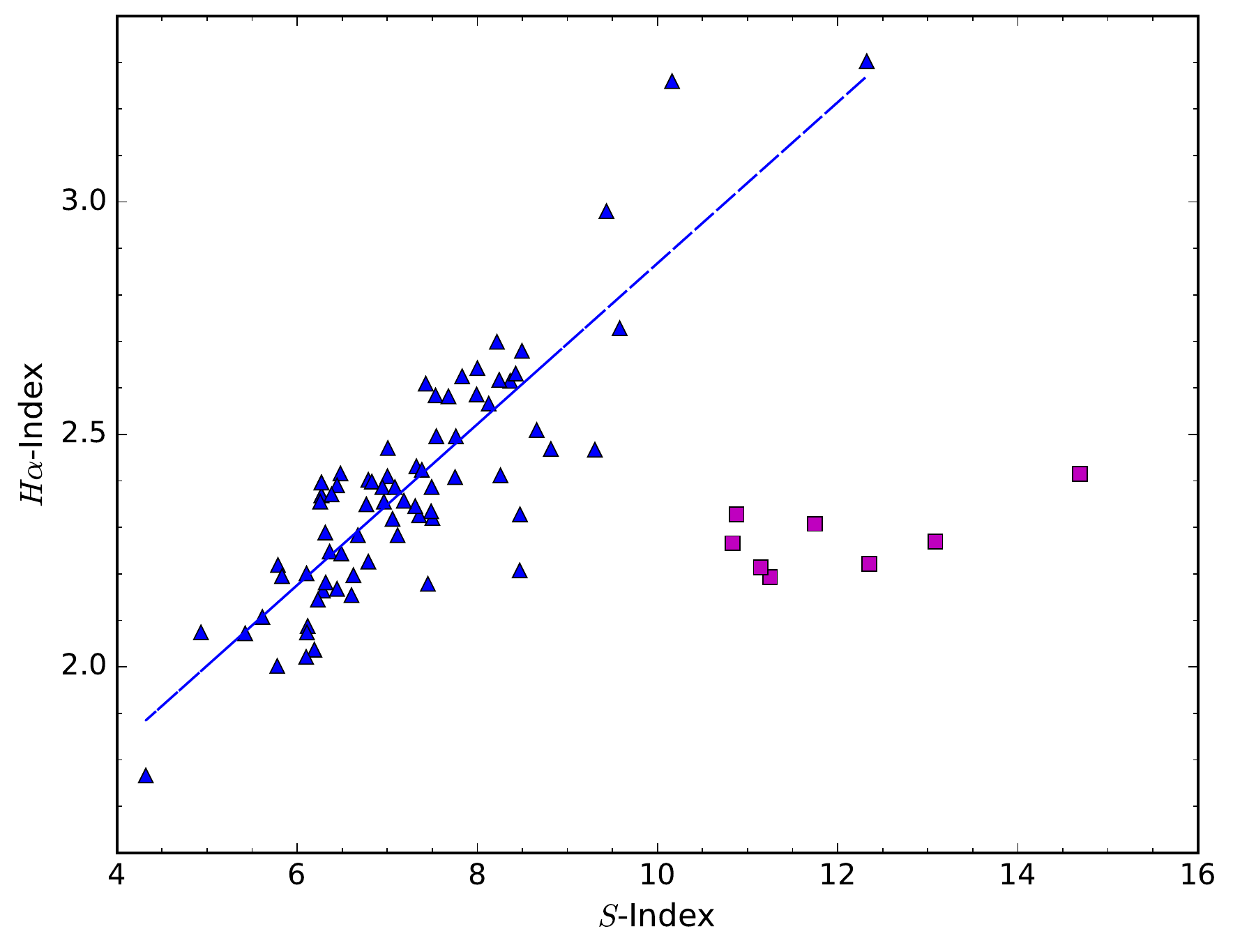}
      \includegraphics[width=9cm]{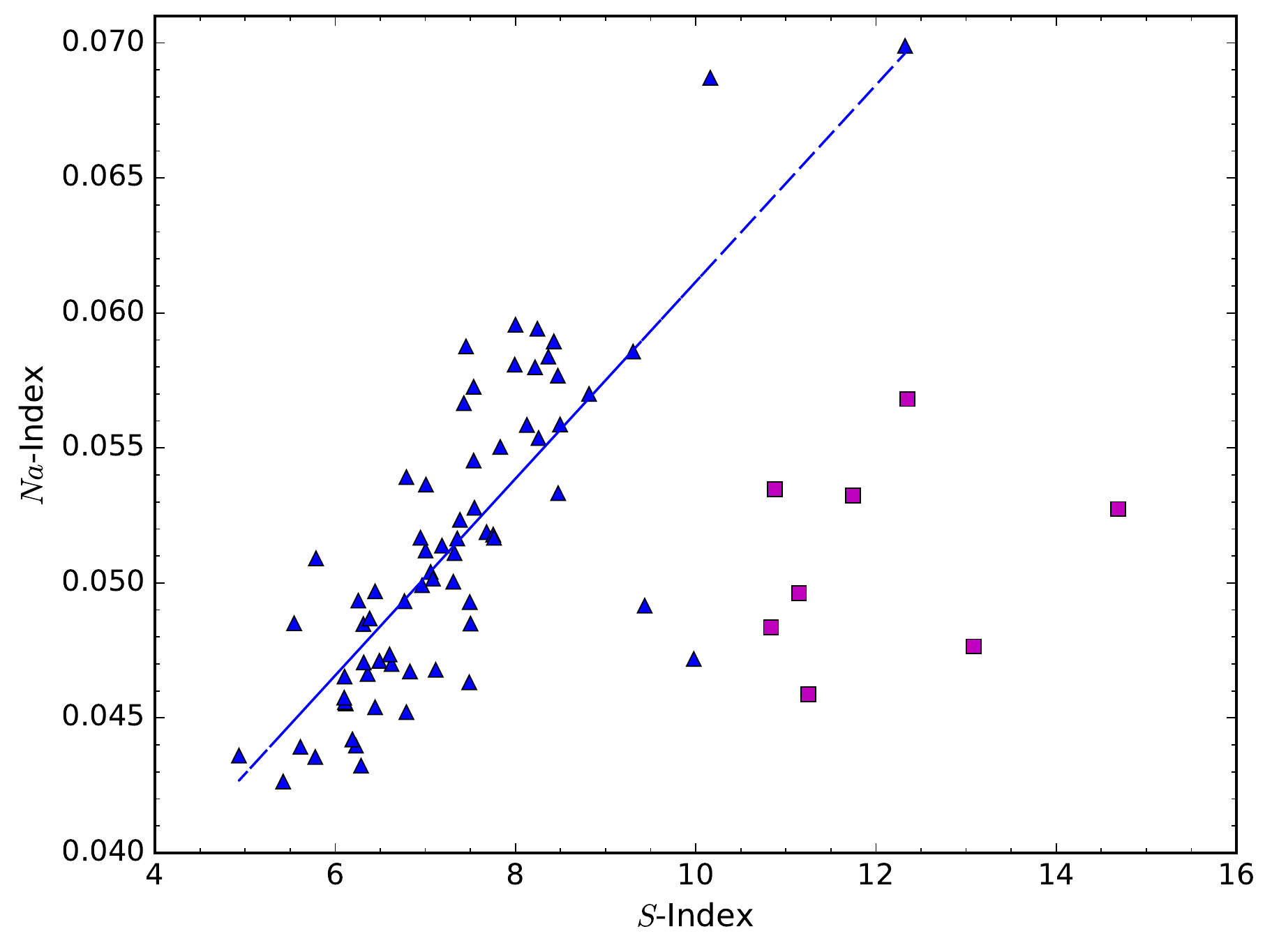}
   \caption{Simultaneous measurements of the H$\alpha$ (\textit{top}) and the Na \scriptsize{I}\normalsize\- (\textit{bottom}) indexes vs. the Mount Wilson $S$-index for spectra that include the two features. The magenta squares indicate the points at which the magnetic activity level increased, some of them associated with flares (gray circles in Fig. \ref{st_CHFUXI}).}%
   \label{indices}
    \end{figure}

The spectra we use in this paper, with the exception of those taken by HIRES and FEROS, cover a wavelength range that allows us to explore different chromospheric features. In particular, we inspect the correlation of  simultaneous measurements of the Ca \scriptsize{II}\normalsize\- lines with the H${\alpha}$ and Na \scriptsize{I}\normalsize\- D lines obtained for 94 and 87 spectra, respectively.
To do so, we computed the sodium Na index and the H${\alpha}$ index as defined by \cite{GomesdaSilva12}
and  \cite{Cincunegui07b}, respectively. The results are shown in Fig. \ref{indices}, and in
the fifth column of Table \ref{obs_table} we indicate the spectra used in this analysis as ``y/y''.

The separation into two groups is clear in both cases. We separated the points with strong magnetic activity marked with gray circles in Fig. \ref{st_CHFUXI}, which are indicated with 
magenta squares. 
For the remaining points, marked in blue, we found strong correlations in both cases, with Pearson coefficients R = 0.86 for H${\alpha}$ and R = 0.79 for Na, as reported by  \cite{Walkowicz09} and \cite{Diaz07b}.
The variability ($\sigma_X / \langle X \rangle $) of the H${\alpha}$ and Na indexes during the cycle (blue points) is about 11 and 12\%, respectively, much smaller than the $S$-index variability, which is about 27\%. 

This strong positive correlation seems to change in the activity cycle phase. The H${\alpha}$ index seems to saturate at a value  $\sim$2.3 at the maximum of the activity cycle (gray points in Fig. \ref{st_CHFUXI}). Furthermore, it is remarkable that this saturation level is lower than the maximum value reached by the H$\alpha$ index.  A similar behavior, but with much more spread, is observed in the Na index.

This saturation might be due to a geometric effect: higher line-fluxes with activity can be due to a larger filling factor of active regions. The area covered by these active regions  increases with height as the magnetic flux tubes spread out, however. While the filling factor increases, these tubes approach each other, until they eventually cover the entire surface higher up in the chromosphere,  at the height where H$\alpha$ and the Na D lines are formed, reaching saturation. 

%__________________________________________________________________
%__________________________________________________________________
%__________________________________________________________________

\section{Conclusions and summary}\label{sec.concl}

New interests in the magnetic activity of M stars  have emerged in the past decades. On the one hand, their strong and moderate flares might severely constrain the habitability of a terrestrial planet (e.g., \citealt{Buccino07,Vida17}). 
On the other hand, due to their internal structures, almost and fully convective stars might be special laboratories for testing the dynamo theory.
Nevertheless, magnetic activity cycles seem to be present in early, mid-, and late-M dwarfs (e.g., \citealt{Cincunegui07, Buccino11, Buccino14, Wargelin17, Ibanez18, Ibanez19,Toledo19}).  
Furthermore, regardless of whether  these stars are fully convective, they are placed indistinctly in the two regimes that are usually observed in the relation of  magnetic activity with the stellar parameters (e.g.,  $L_X/L_{bol}$ vs. $P_{rot}/\tau$ in \citealt{Wright11, Wright18}; $\log R'_{HK}$ vs. $P_{rot}$ in  \citealt{Astudillo17}).
Thus, the change in the internal structure does not affect the activity-rotation relationship. One of our main conclusion is that stellar rotation could drive the activity in partly and fully convective stars. 

We have contributed to this discussion with an exhaustive study of the fast rotator dM4 active star, Gl 729.
First, we studied its short-term variability due to flares and rotation by a detailed analysis of the long-cadence \textit{Kepler} light curve. 
We found that  this active star presents a flare frequency of 0.5 flares per day with energies between $10^{32}$ erg ans $10^{34}$ erg, which implies that energy release during these transient events reaches two or four orders of magnitude more than the quiescent level.
Similar flares were also observed in other M stars (e.g., \citealt{Hawley14, Gunther20}).
After discarding the observations associated with flares in the \textit{Kepler} light curve, we obtained a double-peaked harmonic curve associated with a rotation period of 2.848 days. 

Long-term activity of M stars has been scarcely explored in comparison to solar-type stars.
In this work, we derived the Mount Wilson $S$ index using CASLEO, HARPS, FEROS, UVES, XSHOOTER, and HIRES spectra.
For the whole time series covering 21 years (1998-2019), we detected two significant periods of $\sim 4.2$ yr and $\sim 0.8$ yr with four different tools.
Based on our wavelet analysis in Fig. \ref{wave} we can determine the distribution of the cyclic signal. 
We suspect that the 4.2-year period is modulated by a decadal activity cycle presenting a minimum in the time range xJD=1000-4000 days. 
Altough this behavior has been observed in several solar-type stars (\citealt{Metcalfe13, Radick18}), it has never been reported in M dwarfs.

We also derived a mean activity level of $\log R'_{HK} = -4.645$ in the time span, in agreement with the value reported in \cite{Astudillo17}.  
Nevertheless, given its rotation period,  Gl 729  is placed markedly below the saturation regime in the $\log R'_{HK}$ versus $P_{rot}$ diagram (see Fig. \ref{a-d_diagram}), probably  indicating that its activity could be driven by a  type of dynamo different than for other stars in the graph.

In order to explore the nature of the mechanisms responsible for the cyclical long-term  activity detected in Gl 729, we analyzed its surface differential rotation, which is a necessary ingredient for the $\alpha\Omega$ dynamo operation \citep{Charbonneau10}, whereas it seems not to be determinant  in turbulent $\alpha^2$ dynamos (e.g., \citealt{Durney93,Chabrier06}).
To do so, we studied the \textit{Kepler} light curve with the spot model described in \S \ref{methods} . 
In this analysis, we identified two active longitudes, and we note some indication of an oscillation in the longitude of the dominant center of activity. 
However, the amplitude of this migration, which is about 50$^{\circ}$, is comparable with the longitude resolution achieved by our spot modeling  and it is therefore no evidence of surface differential rotation. 
Thus, the activity cycle of Gl 729 could be driven by a turbulent $\alpha^2$ dynamo. 

\cite{Cole16} studied the oscillatory $\alpha^2$ dynamo for a 1D mean-field dynamo model. They concluded that long activity cycles can be driven by an $\alpha^2$ dynamo under certain conditions if the turbulent diffusivity profile is fairly concentrated toward the equator.

Furthermore, the two activity cycles detected in Gl 729, within the statistical error,  belong  to the active and inactive branch in the $P_{cyc}-P_{rot}$ diagram in \cite{Bohm07}. 
Although this diagram is mainly composed of FGK stars whose activity cycles are probably well reproduced by a solar-type dynamo (e.g., \citealt{Buccino20}), this bimodal distribution in $P_{cyc}-P_{rot}$ could be extended to all cyclic stars, independently of the underlying type of dynamo (see Fig. \ref{b-v_diagram}).
All these facts make Gl 729 an ideal target for an exploration with dynamo models because it shows signs of long-term cyclic magnetic activity, without evidence of surface differential rotation, which is compatible with a nonsolar type dynamo.

\longtab{
\begin{longtable}{ccccccc}
\caption{\label{obs_table}Log of the observations. Col. 1: Spectrograph name; Col. 2: xJD = JD - 2 450 000, where JD is the Julian date; Col. 3: mean Mount Wilson $S$-index; Col. 4: exposure time in seconds; Col. 5: spectrum used in the analysis of section  \S\ref{sec.res} (“y”) or not (“n”), those with a double asterisk were not used in the study because they include flares; Col 6: CASLEO label corresponding to the date of each observing run (MMYY); Col. 7: ID programs of the Gl 729 public observations.} \\
\hline\hline
\multicolumn{1}{l}{Spectrograph} & xJD         & $\langle S \rangle$ & Exp. Time & \multicolumn{1}{l}{$H_{\alpha} / Na I$} & ID           & Program ID                      \\ %\hline
                                 &             &                     & (s)       &                                         & (only REOSC) &                                 \\ 
\hline
\endfirsthead
\caption{continued.} \\
\hline\hline
\multicolumn{1}{l}{Spectrograph} & xJD         & $\langle S \rangle$ & Exp. Time & \multicolumn{1}{l}{$H_{\alpha} / Na I$} & ID           & Program ID                      \\ %\hline
                                 &             &                     & (s)       &                                         & (only REOSC) &                                 \\ 
\hline
\endhead \\
\hline
\endfoot
\multirow{19}{*}{REOSC}          & 3545.000000 & 6.789795e+00        & 8400      & y / y                                   & 0605c        & \multirow{19}{*}{}              \\ %\cline{2-6}
                                 & 3926.000000 & 9.304848e+00        & 5400      & y / y                                   & 0706a        &                                 \\ %\cline{2-6}
                                 & 3955.000000 & 7.451568e+00        & 7200      & y / y                                   & 0806a        &                                 \\ %\cline{2-6}
                                 & 4165.000000 & 7.487149e+00        & 5400      & y / y                                   & 0307c        &                                 \\ %\cline{2-6}
                                 & 4281.000000 & 4.319932e+00        & 7200      & y / n                                   & 0607a        &                                 \\ %\cline{2-6}
                                 & 4367.000000 & 5.787213e+00        & 10800     & y / y                                   & 0907a        &                                 \\ %\cline{2-6}
                                 & 4730.000000 & 1.124911e+01        & 10800     & n / n **                                & 0908b        &                                 \\ %\cline{2-6}
                                 & 4988.000000 & 1.235089e+01        & 10800     & y / y                                   & 0609a        &                                 \\ %\cline{2-6}
                                 & 5106.000000 & 1.087805e+01        & 10800     & y / y                                   & 1009a        &                                 \\ %\cline{2-6}
                                 & 5263.000000 & 1.083528e+01        & 7200      & y / y                                   & 0310c        &                                 \\ %\cline{2-6}
                                 & 5358.000000 & 8.470264e+00        & 10800     & y / y                                   & 0610c        &                                 \\ %\cline{2-6}
                                 & 5731.000000 & 6.674119e+00        & 10800     & y / n                                   & 0611b        &                                 \\ %\cline{2-6}
                                 & 5819.000000 & 9.433980e+00        & 7200      & y / y                                   & 0911a        &                                 \\ %\cline{2-6}
                                 & 6089.000000 & 1.308467e+01        & 5400      & y / y                                   & 0612c        &                                 \\ %\cline{2-6}
                                 & 6172.000000 & 1.174740e+01        & 7200      & y / y                                   & 0912b        &                                 \\ %\cline{2-6}
                                 & 6434.000000 & 1.468871e+01        & 5400      & n / n **                                & 0513a        &                                 \\ %\cline{2-6}
                                 & 6833.000000 & 8.659113e+00        & 5400      & y / n                                   & 0614b        &                                 \\ %\cline{2-6}
                                 & 8273.000000 & 1.114678e+01        & 7200      & y / y                                   & 0618a        &                                 \\ %\cline{2-6}
                                 & 8653.000000 & 9.978682e+00        & 6000      & y / y                                   & 0619d        &                                 \\ \hline
\multirow{10}{*}{HARPS}          & 3574.136719 & 6.789795e+00        & 900       & y / y                                   &              & \multirow{2}{*}{072.C-0488(E)}  \\ %\cline{2-6}
                                 & 3607.078125 & 7.501529e+00        & 900       & y / y                                   &              &                                 \\ \cline{2-7} 
                                 & 7492.320312 & 7.002447e+00        & 900       & y / y                                   &              & \multirow{27}{*}{097.C-0864(B)} \\ %\cline{2-6}
                                 & 7492.328125 & 6.360152e+00        & 900       & y / y                                   &              &                                 \\ %\cline{2-6}
                                 & 7493.335938 & 7.536248e+00        & 900       & n / y                                   &              &                                 \\ %\cline{2-6}
                                 & 7590.027344 & 6.963389e+00        & 900       & y / y                                   &              &                                 \\ %\cline{2-6}
                                 & 7590.039062 & 6.768096e+00        & 900       & y / y                                   &              &                                 \\ %\cline{2-6}
                                 & 7593.023438 & 7.353976e+00        & 900       & y / y                                   &              &                                 \\ %\cline{2-6}
                                 & 7593.035156 & 7.058867e+00        & 900       & y / y                                   &              &                                 \\ %\cline{2-6}
                                 & 7595.117188 & 8.495354e+00        & 1200      & y / y                                   &              &                                 \\ %\cline{2-6}
                                 & 7598.207031 & 7.544929e+00        & 900       & y / y                                   &              &                                 \\ %\cline{2-6}
                                 & 7599.027344 & 8.256663e+00        & 900       & y / y                                   &              &                                 \\ %\cline{2-6}
                                 & 7599.035156 & 8.816503e+00        & 900       & y / y                                   &              &                                 \\ %\cline{2-6}
                                 & 7600.050781 & 7.679464e+00        & 900       & y / y                                   &              &                                 \\ %\cline{2-6}
                                 & 7600.062500 & 6.828854e+00        & 900       & y / y                                   &              &                                 \\ %\cline{2-6}
                                 & 7603.136719 & 7.115284e+00        & 900       & y / y                                   &              &                                 \\ %\cline{2-6}
                                 & 7603.148438 & 6.286373e+00        & 900       & y / y                                   &              &                                 \\ %\cline{2-6}
                                 & 7666.992188 & 8.365159e+00        & 900       & y / y                                   &              &                                 \\ %\cline{2-6}
                                 & 7667.003906 & 8.243643e+00        & 900       & y / y                                   &              &                                 \\ %\cline{2-6}
                                 & 7667.980469 & 6.312413e+00        & 900       & y / y                                   &              &                                 \\ %\cline{2-6}
                                 & 7667.992188 & 7.991933e+00        & 900       & y / y                                   &              &                                 \\ %\cline{2-6}
                                 & 7671.023438 & 6.624882e+00        & 900       & y / y                                   &              &                                 \\ %\cline{2-6}
                                 & 7904.171875 & 6.255995e+00        & 900       & y / y                                   &              &                                 \\ %\cline{2-6}
                                 & 7904.183594 & 6.442607e+00        & 900       & y / y                                   &              &                                 \\ %\cline{2-6}
                                 & 7905.171875 & 5.422745e+00        & 900       & y / y                                   &              &                                 \\ %\cline{2-6}
                                 & 7905.183594 & 5.613698e+00        & 900       & y / y                                   &              &                                 \\ %\cline{2-6}
                                 & 7907.125000 & 8.217604e+00        & 900       & y / y                                   &              &                                 \\ %\cline{2-6}
                                 & 7907.136719 & 7.831357e+00        & 900       & y / y                                   &              &                                 \\ %\cline{2-6}
                                 & 7918.199219 & 8.473655e+00        & 900       & y / y                                   &              &                                 \\ \cline{2-7} 
                                 & 7934.160156 & 7.006788e+00        & 900       & y / y                                   &              & \multirow{31}{*}{099.C-0880(A)} \\ %\cline{2-6}
                                 & 7936.128906 & 7.427752e+00        & 900       & y / y                                   &              &                                 \\ %\cline{2-6}
                                 & 7937.148438 & 7.536248e+00        & 900       & y / y                                   &              &                                 \\ %\cline{2-6}
                                 & 7942.093750 & 6.381851e+00        & 900       & y / y                                   &              &                                 \\ %\cline{2-6}
                                 & 7943.082031 & 5.544260e+00        & 900       & n / y                                   &              &                                 \\ %\cline{2-6}
                                 & 7944.066406 & 8.000612e+00        & 900       & y / y                                   &              &                                 \\ %\cline{2-6}
                                 & 7945.109375 & 6.442607e+00        & 900       & y / y                                   &              &                                 \\ %\cline{2-6}
                                 & 7946.238281 & 7.323596e+00        & 900       & y / y                                   &              &                                 \\ %\cline{2-6}
                                 & 7953.214844 & 7.492851e+00        & 900       & y / y                                   &              &                                 \\ %\cline{2-6}
                                 & 7954.136719 & 1.232309e+01        & 900       & y / y                                   &              &                                 \\ %\cline{2-6}
                                 & 7961.140625 & 6.117120e+00        & 900       & y / y                                   &              &                                 \\ %\cline{2-6}
                                 & 7962.132812 & 8.425916e+00        & 900       & y / y                                   &              &                                 \\ %\cline{2-6}
                                 & 7964.144531 & 6.946029e+00        & 900       & y / y                                   &              &                                 \\ %\cline{2-6}
                                 & 7972.152344 & 6.316752e+00        & 900       & y / y                                   &              &                                 \\ %\cline{2-6}
                                 & 7974.152344 & 6.490346e+00        & 900       & y / y                                   &              &                                 \\ %\cline{2-6}
                                 & 7979.214844 & 8.126467e+00        & 900       & y / y                                   &              &                                 \\ %\cline{2-6}
                                 & 7985.027344 & 7.753242e+00        & 900       & y / y                                   &              &                                 \\ %\cline{2-6}
                                 & 7986.039062 & 7.384354e+00        & 900       & y / y                                   &              &                                 \\ %\cline{2-6}
                                 & 7987.050781 & 6.603183e+00        & 900       & y / y                                   &              &                                 \\ %\cline{2-6}
                                 & 7990.011719 & 7.761920e+00        & 900       & y / y                                   &              &                                 \\ %\cline{2-6}
                                 & 7992.027344 & 5.778612e+00        & 900       & y / y                                   &              &                                 \\ %\cline{2-6}
                                 & 7993.011719 & 6.229956e+00        & 900       & y / y                                   &              &                                 \\ %\cline{2-6}
                                 & 7994.000000 & 7.310576e+00        & 900       & y / y                                   &              &                                 \\ %\cline{2-6}
                                 & 7995.003906 & 9.580315e+00        & 900       & y / n                                   &              &                                 \\ %\cline{2-6}
                                 & 7996.007812 & 1.016185e+01        & 900       & y / y                                   &              &                                 \\ %\cline{2-6}
                                 & 7998.039062 & 6.108440e+00        & 900       & y / y                                   &              &                                 \\ %\cline{2-6}
                                 & 8001.042969 & 7.184721e+00        & 900       & y / y                                   &              &                                 \\ %\cline{2-6}
                                 & 8002.031250 & 7.084904e+00        & 900       & y / y                                   &              &                                 \\ %\cline{2-6}
                                 & 8018.035156 & 6.190897e+00        & 900       & y / y                                   &              &                                 \\ %\cline{2-6}
                                 & 8019.003906 & 6.104101e+00        & 900       & y / y                                   &              &                                 \\ %\cline{2-6}
                                 & 8024.089844 & 6.099760e+00        & 900       & y / y                                   &              &                                 \\ \hline
FEROS                            & 4166.343750 & 7.695313e+00        & 454       & n / n                                   &              & 078.A-9058(A)                   \\ \hline
\multirow{4}{*}{UVES}            & 5716.335938 & 6.269531e+00        & 230       & y / n                                   &              & \multirow{3}{*}{087.D-0069(A)}  \\ %\cline{2-6}
                                 & 5716.335938 & 6.269531e+00        & 230       & y / n                                   &              &                                 \\ %\cline{2-6}
                                 & 5716.339844 & 6.480469e+00        & 230       & y / n                                   &              &                                 \\ \cline{2-7} 
                                 & 7220.070312 & 5.832031e+00        & 500       & y / n                                   &              & 095.D-0685(A)                   \\ \hline
\multirow{2}{*}{XSHOOTER}        & 7185.191406 & 4.931776e+00        & 120       & y / y                                   &              & \multirow{2}{*}{095.D-0949(A)}  \\ \cline{2-6}
                                 & 7185.195312 & 4.783937e+00        & 120       & n / n                                   &              &                                 \\ \hline
\multirow{13}{*}{HIRES}          & 1050.335938 & 4.758584e+00        & 400       & n / n                                   &              & \multirow{2}{*}{U05H}           \\ \cline{2-6}
                                 & 1051.359375 & 4.904892e+00        & 600       & n / n                                   &              &                                 \\ \cline{2-7} 
                                 & 1312.566406 & 4.888838e+00        & 450       & n / n                                   &              & N22H                            \\ %\cline{2-7} 
                                 & 1367.378906 & 4.686143e+00        & 400       & n / n                                   &              & N20H                            \\ %\cline{2-7} 
                                 & 1703.480469 & 4.811812e+00        & 500       & n / n                                   &              & N31H                            \\ %\cline{2-7} 
                                 & 2390.609375 & 7.101417e+00        & 480       & n / n                                   &              & N11H                            \\ %\cline{2-7} 
                                 & 2803.515625 & 4.032948e+00        & 480       & n / n                                   &              & U16H                            \\ %\cline{2-7} 
                                 & 2834.394531 & 4.627389e+00        & 600       & n / n                                   &              & N15H                            \\ %\cline{2-7} 
                                 & 3153.507812 & 4.281481e+00        & 480       & n / n                                   &              & U10H                            \\ %\cline{2-7} 
                                 & 3239.343750 & 5.095374e+00        & 500       & n / n                                   &              & U09H                            \\ %\cline{2-7} 
                                 & 3548.460938 & 6.789794e+00        & 500       & n / n                                   &              & N59H                            \\ %\cline{2-7} 
                                 & 3934.335938 & 6.148695e+00        & 500       & n / n                                   &              & N054Hr                          \\ %\cline{2-7} 
                                 & 5409.390625 & 6.813473e+00        & 600       & n / n                                   &              & H222Hr                          \\ %\hline
                                 &             &                     &           &                                         &              &                                 \\ %\hline
%\multicolumn{1}{l}{}             &             &                     &           & \multicolumn{1}{l}{}                    &              &                                 \\ 
\label{tabla_obs}
\end{longtable}
}

%________________________________________________________________

%\begin{acknowledgements}
%      Part of this work was supported by the German
%      \emph{Deut\-sche For\-schungs\-ge\-mein\-schaft, DFG\/} project
%      number Ts~17/2--1.
%\end{acknowledgements}

%________________________________________________________________

\bibliographystyle{aa}
\small
\bibliography{biblio}
\end{document}